\documentclass[preprint,3p,12pt]{elsarticle}
\usepackage{mathrsfs}
\usepackage{amsmath}
\usepackage{stmaryrd}
\usepackage{bbding}
\usepackage{dcolumn}
\usepackage{graphicx}
\usepackage{amsfonts}
\usepackage{amssymb}
\usepackage{psfrag}
\usepackage{wrapfig}
\usepackage{subfigure}
\usepackage{makeidx}
\usepackage{bm}
\usepackage{epsf}
\usepackage{epsfig}
\usepackage{setspace}
\usepackage{graphicx}
\usepackage{epstopdf}
\usepackage{psfrag}
\usepackage{subfigure}
\usepackage{color}
\begin{document}
	\title{Implicit High-Order Gas Kinetic Scheme for Turbulence Simulation}
	
	\author[HKUST1]{Guiyu Cao}
	\ead{gcaoaa@connect.ust.hk}
	
	\author[NPU]{Hongmin Su}
	\ead{hongminsu@mail.nwpu.edu.cn}
	
	\author[JNS]{Jinxiu Xu}
	\ead{jinxiu3186@sina.com}
	
	\author[HKUST1,HKUST2]{Kun Xu\corref{cor}}
	\ead{makxu@ust.hk} \cortext[cor]{Corresponding author}
	
	\address[HKUST1]{Department of Mathematics, Department of Mechanical and Aerospace Engineering, Hong Kong University of Science and Technology, Clear Water Bay, Kowloon, Hong Kong}
	\address[NPU]{National Key Laboratory of Science and Technology on Aerodynamic Design and Research, \\ Northwestern Polytechnical University, Xi$^{\text{'}}$an, Shaanxi 710072, China}
	\address[JNS]{JiangNan Institute of Computing  Technology, Wuxi, Jiangsu 214121, China}
	\address[HKUST2]{HKUST Shenzhen Research Institute, Shenzhen 518057, China}

\begin{abstract}
	In recent years, coupled with traditional turbulence models, the second-order gas-kinetic scheme (GKS) has been used in the turbulent flow simulations. At the same time, high-order GKS has been developed, such as the two-stage fourth-order scheme (S2O4) GKS, and used for laminar flow calculations. In this paper, targeting on the high-Reynolds number engineering turbulent flows, an implicit high-order GKS with Lower-Upper Symmetric Gauss-Seidel (LU-SGS) technique is developed under the S2O4 framework. Based on Vreman-type LES model and $k - \omega$ SST model, a turbulent relaxation time is obtained and used for an enlarged particle collision time in the implicit high-order GKS for the high-Reynolds number turbulent flows. Numerical experiments include incompressible decaying homogeneous isotropic turbulence, incompressible high-Reynolds number flat plate turbulent flow, incompressible turbulence around NACA0012 airfoil, transonic turbulence around RAE2822 airfoil, and transonic high-Reynolds number ARA M100 wing-body turbulence. Comparisons among the numerical solutions from current implicit high-order GKS, the explicit high-order GKS, the implicit second-order GKS, and experimental measurements have been conducted. Through these examples, it is concluded that the high-order GKS has high accuracy in space and time, especially for smooth flows, obtaining more accurate turbulent flow fields on coarse grids compared with second-order GKS. In addition, significant acceleration on computational efficiency, as well as super robustness in simulating complex flows are confirmed for current implicit high-order GKS. This study also indicates that turbulence modeling plays a dominant role in the capturing physical solution, such as in the transonic three-dimensional complex RANS simulation, in comparison with numerical discretization errors.
\end{abstract}


\maketitle

\section{Introduction}
Turbulence is an important research object among physics, applied mathematics, and engineering applications \cite{tennekes1972first}. Because of its multi-scale features in space and time, it is a challenge to properly balance the accuracy requirements and computational costs \cite{pope2001turbulent} in the simulations, especially for high-Reynolds number turbulent flows. Currently, there are mainly four approaches for turbulence simulation, namely direct numerical simulation (DNS), large eddy simulation (LES), Reynolds averaged Navier-Stokes (RANS), and hybrid RANS/LES methods.

Theoretically,  DNS \cite{kim1987turbulence, moin1998direct, he2018ground} is supposed to resolve turbulent structures above the Kolmogrov dissipation scale \cite{kolmogorov1941local} by grid and time step resolution, but the prohibitive cost limits DNS's engineering applications. In order to study  turbulent flow on unresolved grids, such as for the high-Reynolds number turbulence problems, the RANS models \cite{jones1972prediction, wilcox2008formulation, spalart1992one, menter1994two, menter2003ten, wilcox1993turbulence}, the LES models \cite{manabe1965simulated, germano1991dynamic, vreman2004eddy, sagaut2006large}, and the hybrid RANS/LES methods \cite{frohlich2008hybrid, spalart2009detached, sagaut2013multiscale} have been developed and applied. RANS captures turbulent structures above integral scale under the constraints of computational resources, which has been widely used in engineering turbulence simulations \cite{catalano2003evaluation}. LES solves the filtered Navier-Stokes equations with resolvable turbulent structures above the inertial scale. Even though LES is quite expansive compared with RANS, for unsteady separation turbulent flows, LES has gradually become an indispensable tool to obtain high-resolution turbulent flow fields. To combine the advantages of RANS and LES, the hybrid RANS/LES methods have been proposed and become hot topics in turbulence simulations, which keep good balance between resolution accuracy and computational cost.

In the past decades, the second-order gas-kinetic scheme (GKS) \cite{xu2001gas, kun2014direct} based on the Bhatnagar-Gross-Krook (BGK) \cite{bhatnagar1954model} model has achieved great success for laminar flow computations from incompressible low-speed flow to hypersonic one. It has been extended to flows with multi-temperature \cite{xu2008multiple, cao2018physical}, gravity field \cite{tian2007three}, and magnetohydrodynamics \cite{tang2010gas}. For turbulent flows, GKS can be directly used as a DNS for low-Reynolds number flow \cite{fu2006numerical, kumar2013weno}. The "mixing time" was proposed for kinetic equation based methods for high-Reynolds number turbulence \cite{chen2003extended,chen2004expanded}, which can be regarded as an extension of BGK model with a newly defined collision (relaxation) time $\tau_t$. Following this "mixing time" concept, the second-order gas kinetic schemes  coupled with S-A model \cite{pan2016gas}, $k - \omega$ SST model \cite{jiang2012implicit, righi2016gas, li2016gas}, Vreman-type LES model, and the hybrid RANS/LES methods \cite{tan2018gas} have been developed and implemented in high-Reynolds number turbulent flow simulations.
Most previous work are based on the explicit second-order GKS coupled with traditional turbulence models.
In view of the high-resolution requirement for turbulence simulation, it is fully legitimate to construct high-order GKS (HGKS) coupled with traditional turbulence models.

In recent years, an accurate and robust two-stage fourth-order (S2O4) GKS \cite{li2016two, pan2016efficient, ji2018family} has been developed for laminar flows, which achieves fourth-order accuracy in space and time and shows high efficiency and robustness in the flow simulations with shocks. Focusing on the extension of the scheme to the three-dimensional turbulent flows,
an implicit high-order GKS (IHGKS) is proposed in this paper. On the one hand, the S2O4 GKS framework is used to provide a solid foundation for obtaining high-resolution flow fields in turbulent flow.
On the other hand, LU-SGS method \cite{yoon1986lu, barakos1998implicit} is implemented to overcome the time step barrier in the explicit scheme, and makes the  Courant-Friedrichs-Lewy (CFL) \cite{courant1928partiellen} number large in the three-dimensional high-Reynolds turbulent flows. In what follows, Section 2 presents the construction of this IHGKS under two-stage fourth-order framework. This is followed by the coupling of Vreman-type LES model \cite{vreman2004eddy} and the $k - \omega$ SST \cite{menter1994two} model in the current IHGKS in Section 3. The numerical simulations from incompressible low-speed to transonic three-dimensional complex turbulent flows will be presented in section 4. And the final section is the conclusion and discussion.

\section{Implicit three-dimensional two-stage high-order GKS solver}
\subsection{Three-dimensional finite volume framework based on BGK model}
Based on particle transport and collision, the Boltzmann equation has been constructed for monotonic dilute gas.
The simplification of the Boltzmann equation given by the BGK model has the following form \cite{bhatnagar1954model},
\begin{align}
	\frac{\partial f}{\partial t} + u \frac{\partial f}{\partial x} + v \frac{\partial f}{\partial y}  + w \frac{\partial f}{\partial z}=  \frac{g - f}{\tau},
	\label{boltamann_bgk_eq}
\end{align}
where $f$ is the number density of molecules at position $(x,y,z)$ and particle velocity $(u,v,w)$ at time $t$. The left side of the Eq.(\ref{boltamann_bgk_eq}) denotes the free transport term, and the right hand side represents the collision term. The relation between distribution function $f$ and macroscopic variables, such as mass, momentum, energy, and stress, can be obtained by taking moments in velocity of the gas distribution function.
The collision operator in BGK model shows simple relaxation process from $f$ to a local equilibrium state $g$, with a characteristic time scale $\tau$, which is related to the viscosity and heat conduction coefficients. The local equilibrium state is a Maxwellian distribution,
\begin{align}
	g = \rho (\frac{\lambda}{\pi})^{\frac{K + 3}{2}}  e^{- \lambda [(u - U)^2 + (v - V)^2 + (w - W)^2 + \xi^2]},
	\label{maxwellian}
\end{align}
where $\rho$ is the density, $(U,V,W)$ are the macroscopic fluid velocity in the $x-$,$y-$ and $z-$ directions. Here $\lambda = m/2k_BT$, $m$ is the molecular mass, $k_B$ is the Boltzmann constant, and $T$ is the temperature. For three-dimensional equilibrium diatomic gas, the total number of degrees of freedom in $\xi$ is $K = 2$, which accounts for the two rotational modes $\xi^2 = \xi_1^2 + \xi_2^2$, and the specific heat ratio $\gamma = (K + 5)/(K + 3)$ is determined.

The relation between mass $\rho$, momentum$(\rho U, \rho V, \rho W)$, total energy $\rho E$ with the distribution function $f$ is given by,
\begin{align}
	Q =
	\begin{pmatrix}
		\rho \\
		\rho U \\
		\rho V \\
		\rho W \\
		\rho E
	\end{pmatrix}
	= \int \psi_{\alpha} f d \Xi, \ \ \alpha = 1, 2, 3, 4, 5,
\end{align}
where $d \Xi = du dv dw d\xi_1 d\xi_2$ and $\psi_{\alpha}$ is the component of the vector of collision invariants
\begin{align*} \psi = (\psi_1, \psi_2, \psi_3, \psi_4, \psi_5)^T = (1, u, v, w, \frac{1}{2}(u^2 + v^2 + w^2 + \xi^2))^T.
\end{align*}
Since only mass, momentum and total energy are conserved during particle collisions, the compatibility condition for the collision term turns into,
\begin{align}
	\int \frac{g - f}{\tau} \psi d \Xi = 0,
\end{align}
at any point in space and time.

Based on the above BGK model as Eq.(\ref{boltamann_bgk_eq}), the Euler equations can be obtained for a local equilibrium state with $f = g$. On the other hand, the Navier-Stokes equations, the stress and Fourier heat conduction terms can be derived with the Chapman-Enskog expansion \cite{chapman1970mathematical} truncated to the $1$st-order of $\tau$,
\begin{align}
	f = g + Kn f_1 = g - \tau (\frac{\partial{g}}{\partial t} + u \frac{\partial g}{\partial x} + v \frac{\partial g}{\partial y} + w \frac{\partial g}{\partial z}) .
	\label{ce_expansion}
\end{align}
For the Burnett and super-Burnett solutions, the above expansion can be naturally extended \cite{ohwada2004kinetic}, such as $f = g + Kn f_1 + Kn^2 f_2 + Kn^3 f_3 + \cdots$. For the above Navier-Stokes solutions, the GKS based on the kinetic BGK model has been well developed \cite{xu2001gas}. In order to simulate the flow with any realistic Prandtl number, a modification of the heat flux in the energy transport is used in this scheme, which is also implemented in the present study.

Taking moments of Eq.(\ref{boltamann_bgk_eq}) and integrating over the control volume $\Omega_{ijk} = \overline{x_i} \times \overline{y_j} \times \overline{z_k}$ with $\overline{x_i} = [x_i - \frac{\Delta x}{2}, x_i + \frac{\Delta x}{2}]$, $\overline{y_j} = [y_j - \frac{\Delta y}{2}, y_j + \frac{\Delta y}{2}]$, $\overline{z_k} = [z_k - \frac{\Delta z}{2}, z_k + \frac{\Delta z}{2}]$, the three-dimensional finite volume scheme can be written as
\begin{equation}
	\label{semi_scheme}
	\begin{aligned}
		&\frac{d Q_{ijk}}{d t} = \mathbf{L}(Q_{ijk}) = \frac{1}{|\Omega_{ijk}|} [\int_{\overline{y_j} \times \overline{z_k}} (F_{i - 1/2, j, k} - F_{i + 1/2, j, k}) dy dz \\
		&+ \int_{\overline{x_i} \times \overline{z_k}} (G_{i, j - 1/2, k} - G_{i, j + 1/2, k}) dx dz + \int_{\overline{x_i} \times \overline{y_j}} (H_{i, j, k - 1/2} - H_{i, j, k + 1/2}) dx dy],
	\end{aligned}
\end{equation}
where $Q_{ijk}$ are the cell averaged conservative flow variables, i.e., mass, momentum and total energy. All of them are averaged over control volume $\Omega_{ijk}$ and volume of the numerical cell is $|\Omega_{ijk}| = \Delta x \Delta y \Delta z$. Here, numerical fluxes in $x-\text{direction}$ is presented as an example
\begin{align}
	\int_{\overline{y_j} \times \overline{z_k}}  F_{i + 1/2, j, k} dy dz = F_{\textbf{x}_{i + 1/2, j, k}, t} \Delta y \Delta z.
\end{align}
Based on the fifth-order WENO-JS  spatial reconstruction on the primitive flow variables \cite{jiang1996efficient}, the reconstructed point value and the spatial derivatives in one normal and two tangential directions can be obtained. In the smooth flow computation, the linear form of WENO-JS is adopted to reduce the dissipation. Gaussian points are widely used for high-order finite volume scheme, however, it is very expensive because additional heavy reconstructions and  flux calculations are required at each interface \cite{pan2016efficient}. To save computational resources for three-dimensional high-Reynolds number engineering turbulence problems, Gaussian points have not been used in the IHGKS. The numerical fluxes $F_{\textbf{x}_{i + 1/2, j, k}, t}$ can be provided by the flow solvers, which can be evaluated by taking moments of the gas distribution function as
\begin{align}
	F_{\textbf{x}_{i + 1/2, j, k}, t}= \int \psi_{\alpha} u f(\textbf{x}_{i + 1/2, j, k}, t, \textbf{u}, \xi) d \Xi, \ \ \alpha = 1, 2, 3, 4, 5.
	\label{flux_integration}
\end{align}
Here $f(\textbf{x}_{i + 1/2, j, k}, t, \textbf{u}, \xi)$ is based on the integral solution of BGK equation Eq.(\ref{boltamann_bgk_eq}) at the cell interface
\begin{align}
	f(\textbf{x}_{i + 1/2, j, k}, t, \textbf{u}, \xi_r) = \frac{1}{\tau} \int_0^t g(\textbf{x}', t', \textbf{u}, \xi_r) e^{-(t - t')/\tau} d t' + e^{-t/\tau} f_0(-\textbf{u}t, \xi_r),
\end{align}
where $\textbf{x}_{i + 1/2, j, k} = \textbf{0}$ is the location of cell interface, $\textbf{u} = (u, v, w)$ is the particle velocity, $\textbf{x}_{i + 1/2, j, k} = \textbf{x}' + \textbf{u} (t - t')$ is the trajectory of particles.  $f_0$ is the initial gas distribution, and $g$ is the corresponding intermediate equilibrium state as Eq.(\ref{maxwellian}). $g$ and $f_0$ can be constructed as
\begin{equation*}
	\begin{aligned}
		g = g_0(1 + \overline{a} x + \overline{b} y + \overline{c} z + \overline{A} t),
	\end{aligned}
\end{equation*}
and
\begin{equation*}
	\begin{aligned}
		f_0 =
		\begin{cases}
			g_l [1 +  (a_l x + b_l y + c_l z) - \tau (a_l u + b_l v + c_l w + A_l)], &x \leq 0, \\
			g_r [1 +  (a_r x + b_r y + c_r z) - \tau (a_r u + b_r v + c_r w + A_r)], &x > 0,
		\end{cases}
	\end{aligned}
\end{equation*}
where $g_l$ and $g_r$ are the initial gas distribution functions on both sides of a cell interface. $g_0$ is the initial intermediate equilibrium state located at cell interface, which can be determined through the compatibility condition
\begin{align*}
	\int \psi_{\alpha} g_0 d \Xi = \int_{u>0} \psi_{\alpha} g_l d \Xi + \int_{u<0} \psi_{\alpha} g_r d \Xi, \ \ \alpha = 1, 2, 3, 4, 5.
\end{align*}
For the second-order flux, the time-dependent gas distribution function at cell interfaces is evaluated as
\begin{equation}
	\begin{aligned}
		\label{formalsolution_neq}
		f(\textbf{x}_{i + 1/2, j, k}, t, \textbf{u}, \xi_r) &= (1 - e^{-t/\tau}) g_0 + ((t + \tau) e^{-t\tau} - \tau) (\overline{a} u + \overline{b} v + \overline{c} w) g_0   \\
		&+ (t - \tau + \tau e^{-t\tau}) \overline{A} g_0  \\
		&+ e^{-t/\tau} g_l [1 - (\tau + t) (a_l u + b_l v + c_l w) - \tau A_l] (1 - H(u)) \\
		&+ e^{-t/\tau} g_r [1 - (\tau + t) (a_r u + b_r v + c_r w) - \tau A_r] H(u),
	\end{aligned}
\end{equation}
where the coefficients in Eq.(\ref{formalsolution_neq}) can be determined by the spatial derivatives of macroscopic flow variables and the compatibility condition \cite{kun2014direct}. In smooth flow region, the discontinuities of flow variables at a cell interface disappear, and the gas distribution function at a cell interface $f(\textbf{x}_{i + 1/2, j, k}, t, \textbf{u}, \xi_r)$ automatically reduces to
\begin{equation}
	\begin{aligned}
		\label{formalsolution_smooth}
		f(\textbf{x}_{i + 1/2, j, k}, t, \textbf{u}, \xi_r) &=  g_0 [1 - \tau(\overline{a} u + \overline{b} v + \overline{c} w + \overline{A}) + t\overline{A}].
	\end{aligned}
\end{equation}
The full flux Eq.(\ref{formalsolution_neq}) is necessary for shock-capturing, which must be used for transonic to supersonic flows. While, Eq.(\ref{formalsolution_smooth}) is applied in smooth flows, and the computational costs can be reduced. 

Here, the second-order accuracy in time can be achieved by one step integration from the second-order gas-kinetic solver Eq.(\ref{formalsolution_neq}). Based on a high-order expansion of the equilibrium state around a cell interface, a one-stage third-order GKS has been developed successfully \cite{li2010high, ren2015multi, pan2016third}. However, the one-stage gas-kinetic solver become very complicated, especially for three-dimensional computations.

\subsection{Two-stage high-order temporal discretization}
In recent study, a two-stage fourth-order time-accurate discretization has been developed for Lax-Wendroff flow solvers, particularly applied for hyperbolic equations with the generalized Riemann problem (GRP) solver \cite{li2016two} and the GKS \cite{pan2016efficient}. Such method provides a reliable framework to develop the three-dimensional IHGKS with a second-order flux function Eq.(\ref{formalsolution_neq}) or Eq.(\ref{formalsolution_smooth}). The key point for this two-stage fourth-order method is to use time derivative of flux function.
In order to obtain the time derivative of flux function at $t_n$ and $t_{\ast} = t_n + \Delta t/2$, the flux function should be approximated as a linear function of time within a time interval.

According to the numerical fluxes at cell interface Eq.(\ref{flux_integration}), the following notation is introduced
\begin{align}
	\mathbb{F}_{i + 1/2, j, k}(Q^n, \zeta) &= \int_{t_n}^{t_n + \zeta} \mathbf{F}_{i + 1/2, j, k}(Q^n, t) dt = \int_{t_n}^{t_n + \zeta} F_{\textbf{x}_{i + 1/2, j, k}, t} dt.
	\label{Flux_twostage}
\end{align}
In the time interval $[ t_n, t_n + \Delta t/2]$, the flux is expanded as the following linear form
\begin{align}
	\mathbf{F}_{i + 1/2, j, k} (Q^n, t) = \mathbf{F} (Q^n, t_n) _{i + 1/2, j, k} + \partial_t \mathbf{F} (Q^n, t_n) _{i + 1/2, j, k} (t - t_n).
	\label{linear_flux}
\end{align}
Based on Eq.(\ref{Flux_twostage}) and linear expansion of flux as Eq.(\ref{linear_flux}), the coefficients $\mathbf{F}_{i + 1/2, j, k}(Q^n, t_n)$ and $\partial_t \mathbf{F}_{i + 1/2, j, k}(Q^n, t_n)$ can be determined as,
\begin{align*}
	\mathbf{F}_{i + 1/2, j, k}(Q^n, t_n) \Delta t + \frac{1}{2} \partial_t \mathbf{F}_{i + 1/2, j, k}(Q^n, t_n)  \Delta t^2 &= 	\mathbb{F}_{i + 1/2, j, k}(Q^n, \Delta t), \\
	\frac{1}{2} \mathbf{F}_{i + 1/2, j, k}(Q^n, t_n) \Delta t + \frac{1}{8} \partial_t \mathbf{F}_{i + 1/2, j, k}(Q^n, t_n)  \Delta t^2 &= \mathbb{F}_{i + 1/2, j, k}(Q^n, \Delta t/2).
\end{align*}
By solving the linear system, we have
\begin{equation}
	\begin{aligned}
		\label{flux_der}
		\mathbf{F}_{i + 1/2, j, k}(Q^n, t_n) &= (4 \mathbb{F}_{i + 1/2, j, k}(Q^n, \Delta t/2) - \mathbb{F}_{i + 1/2, j, k}(Q^n, \Delta t))/\Delta t,  \\
		\partial_t \mathbf{F}_{i + 1/2, j, k}(Q^n, t_n) &= 4(\mathbb{F}_{i + 1/2, j, k}(Q^n, \Delta t) - 2\mathbb{F}_{i + 1/2, j, k}(Q^n, \Delta t/2))/\Delta t^2 ,
	\end{aligned}
\end{equation}
and $\mathbf{F}_{i + 1/2, j, k}(Q^{\ast}, t_{\ast})$,$\partial_t \mathbf{F}_{i + 1/2, j, k}(Q^{\ast}, t_{\ast})$ for the intermediate state $t_{\ast}$ can be constructed similarly.

With these notations, the two-stage high-order algorithm for three-dimensional flow is given by the following steps. \\
(i) With the initial reconstruction, update $Q_{ijk}^{\ast}$ at $t_{\ast} = t_n + \Delta t/2$ by
\begin{equation}
	\begin{aligned}
		\label{qq_star}
		Q_{ijk}^{\ast} - Q_{ijk}^{n} = & - \frac{1}{\Delta x} [\mathbb{F}_{i + 1/2, j, k}(Q^n, \Delta t/2) - \mathbb{F}_{i - 1/2, j, k}(Q^n, \Delta t/2)]  \\
		& - \frac{1}{\Delta y} [\mathbb{G}_{i, j + 1/2, k}(Q^n, \Delta t/2) - \mathbb{G}_{i, j - 1/2, k}(Q^n, \Delta t/2)] \\
		& - \frac{1}{\Delta z} [\mathbb{H}_{i, j, k + 1/2}(Q^n, \Delta t/2) - \mathbb{H}_{i, j, k - 1/2}(Q^n, \Delta t/2)],
	\end{aligned}
\end{equation}
and compute the fluxes and their derivatives by Eq.(\ref{flux_der}) for future using,
\begin{align*}
	\mathbf{F}_{i + 1/2, j, k}(Q^n, t_n), \ &\mathbf{G}_{i, j + 1/2, k}(Q^n, t_n), \ \mathbf{H}_{i, j, k + 1/2}(Q^n, t_n), \\
	\partial_t \mathbf{F}_{i + 1/2, j, k}(Q^n, t_n), \ &\partial_t \mathbf{G}_{i, j + 1/2, k}(Q^n, t_n), \ \partial_t \mathbf{H}_{i, j, k + 1/2}(Q^n, t_n).
\end{align*}
(ii) Reconstruct intermediate value $W_{ijk}^{\ast}$ and compute
\begin{align*}
	\partial_t \mathbf{F}_{i + 1/2, j, k}(Q^{\ast}, t_{\ast}), \ &\partial_t \mathbf{G}_{i, j + 1/2, k}(Q^{\ast}, t_{\ast}),\ \partial_t \mathbf{H}_{i, j, k + 1/2}(Q^{\ast}, t_{\ast}),
\end{align*}
where the derivatives are determined by Eq.(\ref{flux_der}) in the time interval $[t_{\ast}, t_{\ast} + \Delta t]$. \\
(iii) Update $Q_{ijk}^{n + 1}$ by
\begin{equation}
	\begin{aligned}
		\label{qq_nn}
		Q_{ijk}^{n + 1} - Q_{ijk}^{n} = &- \frac{\Delta t}{\Delta x}[\overline{\mathbb{F}}^n_{i + 1/2, j, k} - \overline{\mathbb{F}}^n_{i - 1/2, j, k}]  \\
		&- \frac{\Delta t}{\Delta y}[\overline{\mathbb{G}}^n_{i, j + 1/2, k} - \overline{\mathbb{G}}^n_{i, j - 1/2, k}] \\
		&- \frac{\Delta t}{\Delta z}[\overline{\mathbb{H}}^n_{i, j, k + 1/2} - \overline{\mathbb{H}}^n_{i, j, k - 1/2}],
	\end{aligned}
\end{equation}
where $\overline{\mathbb{F}}^n_{i + 1/2, j, k}$, $\overline{\mathbb{G}}^n_{i, j + 1/2, k}$ and $\overline{\mathbb{H}}^n_{i, j, k + 1/2}$ are the numerical fluxes and expressed as
\begin{align*}
	\overline{\mathbb{F}}^n_{i + 1/2, j, k} &= \mathbf{F}_{i + 1/2, j, k}(Q^n, t_n) + \frac{\Delta t}{6} [\partial_t \mathbf{F}_{i + 1/2, j, k}(Q^n, t_n) + 2 \partial_t \mathbf{F}_{i + 1/2, j, k}(Q^{\ast}, t_{\ast})], \\
	\overline{\mathbb{G}}^n_{i, j + 1/2, k} &= \mathbf{G}_{i, j + 1/2, k}(Q^n, t_n) + \frac{\Delta t}{6} [\partial_t \mathbf{G}_{i, j + 1/2, k}(Q^n, t_n) + 2 \partial_t \mathbf{G}_{i, j + 1/2, k}(Q^{\ast}, t_{\ast})], \\
	\overline{\mathbb{H}}^n_{i, j, k + 1/2} &= \mathbf{H}_{i, j, k + 1/2}(Q^n, t_n) + \frac{\Delta t}{6} [\partial_t \mathbf{H}_{i, j, k + 1/2}(Q^n, t_n) + 2 \partial_t \mathbf{H}_{i, j, k + 1/2}(Q^{\ast}, t_{\ast})].
\end{align*}

In summary, with the initial reconstruction, the intermediate state $Q_{ijk}^{\ast}$ is updated by
\begin{equation}
\begin{aligned}
\label{qq_star_simplify}
    Q_{ijk}^{\ast} = Q_{ijk}^{n} + \frac{\Delta t}{2} \mathbf{L}^{\ast} (Q^n), 
\end{aligned}
\end{equation}
where $\frac{\Delta t}{2} \mathbf{L}^{\ast} (Q^n)$ represents the right-hand side of Eq.(\ref{qq_star}). Then, with the prepared fluxes and their derivatives, $Q_{ijk}^{n + 1}$ can be updated by
\begin{equation}
\begin{aligned}
\label{qq_nn_simplify}
Q_{ijk}^{n + 1} =  Q_{ijk}^{n} + \Delta t \mathbf{L}^{n + 1} (Q^n),
\end{aligned}
\end{equation}
where $\Delta t \mathbf{L}^{n + 1} (Q^n)$ is the right-hand side terms in Eq.(\ref{qq_nn}). $\mathbf{L}$ is the spatial discretization operator as Eq.(\ref{semi_scheme}).

\subsection{Implicit LU-SGS method}
In previous work, LU-SGS method has been applied in GKS for hypersonic flows \cite{xu2005multidimensional} and near-continuum flows \cite{li2006applications} in two-dimensional cases. For three-dimensional flow, in order to use large CFL number to increase the computational efficiency, instead of updating  conservative variables explicitly, implicit LU-SGS method is used to update conservative variables $Q_{ijk}^{\ast}$ and $Q_{ijk}^{n + 1}$. As an example, in the following we present the brief updating procedure of intermediate variables $Q^{\ast}$.

Firstly, introduce the Jacobian matrices $\mathcal{A} = \frac{\partial F }{\partial Q}_i$, $\mathcal{B} = \frac{\partial F}{\partial Q}_j$, and $\mathcal{C} = \frac{\partial F}{\partial Q}_k$, with the Euler flux $F$ for laminar flow \cite{yoon1986lu} and the extended flux when coupled with turbulence model \cite{barakos1998implicit}. Based on the LU-SGS technique, Eq.(\ref{qq_star_simplify}) can be written as
\begin{equation}
\begin{aligned}\label{lusgs_qstar}
    &\mathbf{R}^{\ast}(Q^n) = \frac{\Delta t}{2} \mathbf{L}^{\ast} (Q^n), \\
	(L + D)D^{-1}&(D + U) \Delta Q = \mathbf{R}^{\ast}(Q^n),
\end{aligned}
\end{equation}
where $\Delta Q = Q^{\ast} - Q^{n}$, with the matrices $ L = -(\mathcal{A}_{i - 1}^{+} + \mathcal{B}_{j - 1}^{+} + \mathcal{C}_{k - 1}^{+}) $, $ D = \frac{\mathcal{I}}{\Delta t} + \mathcal{A}_{i}^{+} - \mathcal{A}_{i}^{-} + \mathcal{B}_{j}^{+} - \mathcal{B}_{j}^{-} + \mathcal{C}_{k}^{+} - \mathcal{C}_{k}^{-} $, and $ U = \mathcal{A}_{i + 1}^{-} + \mathcal{B}_{j + 1}^{-} + \mathcal{C}_{k + 1}^{-}$. Unknown matrices are introduced by $ \mathcal{A}^{\pm} = \frac{1}{2} (\mathcal{A} \pm r_{\mathcal{A}} \mathcal{I}), r_{\mathcal{A}} = \beta \sigma_{\mathcal{A}}$, $ \mathcal{B}^{\pm} = \frac{1}{2} (\mathcal{B} \pm r_{\mathcal{B}} \mathcal{I}), r_{\mathcal{B}} = \beta \sigma_{\mathcal{B}}$, and $\mathcal{C}^{\pm} = \frac{1}{2} (\mathcal{C} \pm r_{\mathcal{C}} \mathcal{I}), r_{\mathcal{C}} = \beta \sigma_{\mathcal{C}}$. Where $\mathcal{I}$ is the unit matrix, $(\sigma_{\mathcal{A}}, \sigma_{\mathcal{B}}, \sigma_{\mathcal{C}})$ are the spectral radii of the Jacobian matrices, with the coefficient $\beta \ge 1$ to ensure dominant diagonal.

Then, use two-step sweeping way to get the solution $\Delta Q$
\begin{equation}
	\begin{aligned}
		(L + D) \Delta Q^{\circ} &= \mathbf{R}^{\ast}(Q^n), \\
		(D + U) \Delta Q &= D \Delta Q^{\circ}.
	\end{aligned}
\end{equation}
Subsequently, the intermediate macroscopic flow variables $Q^{\ast}$ are updated by
\begin{align}\label{lusge_update}
	Q^{\ast} = Q^n + \Delta Q.
\end{align}

Based on  Eq.(\ref{qq_nn_simplify}), the residual for $t_{n + 1}$ step is defined as $\mathbf{R}^{n + 1}(Q^n) = \Delta t \mathbf{L}^{n + 1} (Q^n)$. Then, similar procedures as Eq.(\ref{lusgs_qstar})-Eq.(\ref{lusge_update}) can be used to update the $t_{n + 1}$ step macroscopic flow variables $Q^{n + 1}$. In this way, within the S2O4 GKS framework, LU-SGS method is implemented to overcome the time step barrier in the explicit scheme.

\section{IHGKS coupled with turbulence model}
We follow the concept of turbulent eddy viscosity \cite{boussinesq1870essai}, which models the effect of unresolved turbulent scales by enlarged turbulent eddy viscosity in turbulence region. Similarly, the enlarging turbulent relaxation time $\tau_t$ is proposed to describe the turbulent flows under the kinetic  framework. Based on this enlarging turbulent relaxation time $\tau_t$ \cite{chen2003extended}, extended BGK model for turbulent flows can be written as,
\begin{align}
	\frac{\partial f}{\partial t} + u \frac{\partial f}{\partial x} + v \frac{\partial f}{\partial y}  + w \frac{\partial f}{\partial z}=  \frac{g - f}{\tau + \tau_{t}}.
	\label{boltamann_bgk_tur}
\end{align}
Using Chapman-Enskog expansion \cite{chen2004expanded}, Eq.(\ref{boltamann_bgk_tur}) can recover traditional RANS turbulent  eddy  viscosity model through the relation between turbulent eddy viscosity $\mu_t$ and turbulent relaxation time $\tau_t$, with
\begin{align}
	\label{mut_taut}
	\tau + \tau_{t} = \frac{\mu + \mu_t}{p},
\end{align}
where $p$ is the pressure. The key point is to get turbulent eddy viscoisity $\mu_t$, then turbulent relaxation time $\tau_{t}$ will be
determined by Eq.(\ref{mut_taut}). In original study \cite{chen2004expanded}, this enlarged relaxation time $\tau_t$ is called "mixing time", which is comparable with the classical concept of "mixing length". In this paper, based on extended BGK model and "mixing time" concept, time-relaxation turbulence simulation will be studied.

In present work, Vreman-type model for LES and $k-\omega$ SST model for RANS simulation will be used to evaluate $\tau_t$ and use the relaxation time $\tau + \tau_{t}$ in Eq.(\ref{boltamann_bgk_tur}).
All conserved macroscopic variables are calculated from GKS, and the turbulent viscosity is obtained from the LES/RANS eddy viscosity model.
The evolution of turbulent variables depends on the conserved macroscopic variables. This coupling process is applied at each step for turbulence simulations.

\subsection{LES: Vreman-type model}
To keep the simple eddy viscosity closure form and overcome the drawbacks of the original Smagorinsky  model \cite{manabe1965simulated}, Vreman-type model \cite{vreman2004eddy} is proposed by A.W. Vreman in a simple algebra form, which is comparable to dynamic Smagorinsky model \cite{germano1991dynamic}. For Vreman-type model, turbulent eddy viscosity $\mu_t$ is given by
\begin{align}
	\label{vreman_les}
	\mu_t &= \rho c \sqrt{\frac{B_{\beta}}{a_{ij} a_{ij}}},
\end{align}
where $\rho$ is the density, and constant $c$ is related to Smagorinsky constant $c = 2.5C_s^2$, with $C_s = 0.1$. Left unknowns in Eq.(\ref{vreman_les}) can be determined through the combination of velocity gradient in resolved flow fields, as
\begin{align}
	\label{vreman_detail}
	\begin{cases}
		\alpha_{ij} &= \frac{\partial U_j}{\partial x_i}, \\
		\beta_{ij}  &= \Delta^2 \alpha_{mi} \alpha_{mj}, \\
		B_{\beta}   &= \beta_{11}\beta_{22}  - \beta_{12}^2 + \beta_{11}\beta_{33} - \beta_{13}^2 +  \beta_{22}\beta_{33} - \beta_{23}^2.
	\end{cases}
\end{align}
In Eq.(\ref{vreman_detail}), the $\frac{\partial U_j}{\partial x_i}$ represents the first-order derivative of cell averaged velocity, and $\Delta$ is the width of the numerical cell.  For averaging process, numerical cell itself acts as the filter and no explicit filter is adopted in current scheme. 

\subsection{RANS: $k-\omega$ SST model}
$k-\omega$ SST model \cite{menter1994two} combines the positive features of $k-\omega$ model \cite{wilcox2008formulation} and $k-\epsilon$ model \cite{jones1972prediction} together. For this model, evolution equation of turbulence kinetic energy $k$ and specific dissipation rate $\omega$ are modeled as
\begin{equation}
\label{k_w_sst}
	\begin{aligned}
		\frac{\partial (\rho k)}{\partial t} + \frac{\partial}{\partial x_j} [\rho U_j k - (\mu + \sigma_k \mu_t) \frac{\partial k}{\partial x_j}] &= P - \beta^{\ast} \rho \omega k, \\
		\frac{\partial (\rho \omega)}{\partial t} + \frac{\partial}{\partial x_j} [\rho u_j \omega - (\mu + \sigma_{\omega} \mu_t) \frac{\partial \omega}{\partial x_j}] &= \frac{\gamma}{\nu_t}P - \beta \rho \omega^2 + 2(1 - F_1) \frac{\rho \sigma_{\omega 2}}{\omega} \frac{\partial k}{\partial x_j} \frac{\partial \omega}{\partial x_j},
	\end{aligned}
\end{equation}
where $P$ is the production of turbulence kinetic energy. In current study, $P$ is written in SST-V2003 form  \cite{menter2003ten}, as
\begin{align*}
	P^{\ast} = \mu_t \Omega^2 - \frac{2}{3} \rho k \delta_{ij} \frac{\partial U_i}{\partial x_j}, \\
	P = min(P^{\ast}, 10 \beta^{\ast} \rho \omega k),
\end{align*}
where $\Omega = \sqrt{\Omega_{ij} \Omega_{ij}}$ is the vorticity magnitude.
The turbulent eddy viscosity is computed from
\begin{align}
	\mu_t &= \frac{ \rho a_1 k}{max\{a_1 \omega, S F_2\}} ,
\end{align}
where $\nu_t = \mu_t /\rho$ is the turbulent kinematic viscosity, $S = \sqrt{2 S_{ij} S_{ij}}$ is the shear strain rate magnitude. $\Omega_{ij}$ and $S_{ij}$ are denoted by
\begin{align*}
	\Omega_{ij} = \frac{1}{2}(\frac{\partial U_i}{\partial x_j} - \frac{\partial U_j}{\partial x_i}), \ S_{ij} = \frac{1}{2}(\frac{\partial U_i}{\partial x_j} + \frac{\partial U_j}{\partial x_i}).
\end{align*}

Each of the constants is a blend of an inner constant and outer constant via
\begin{align*}
	\phi = F_1 \phi_1 + (1 - F_1) \phi_2, \quad (\phi = {\sigma_k, \sigma_{\omega}, \beta, \gamma})
\end{align*}
where $\phi_1$ represents the inner constants of $k-\omega$ model and $\phi_2$ represents the outer constants of the $k-\epsilon$ model. For inner layer,
\begin{align*}
	\sigma_{k1} = 0.85, \ \sigma_{\omega 1} = 0.5, \ \beta_1 = 0.075, \ \gamma_1 = \frac{5}{9}, 
\end{align*}
and for outer layer,
\begin{align*}
	\sigma_{k2} = 1.0, \ \sigma_{\omega 2} = 0.856, \ \beta_2 = 0.0828, \ \gamma_2 = 0.44. 
\end{align*}
$F_1$ and $F_2$ are hybrid functions are given by
\begin{align*}
	F_1 &= \tanh\{min[max(\frac{\sqrt{k}}{\beta^{\ast} \omega d}, \frac{500 \mu}{\rho \omega d^2}), \frac{4 \rho \sigma_{\omega 2} k}{CD_{k\omega} d^2}]\}^4, \\
	F_2 &= \tanh[max(\frac{2 \sqrt{k}}{0.09 \omega d}, \frac{500 \mu}{\rho \omega d^2})]^2, \\
	CD_{k\omega}   &= max(\frac{2 \rho \sigma_{\omega 2}}{\omega} \frac{\partial k}{\partial x_j} \frac{\partial \omega}{\partial x_j}, 10^{-10}),
\end{align*}
where $d$ is the colsest distance from the field point to the nearest wall, and left constants are $a_1 = 0.31$ and $\beta^{\ast} = 0.09$.

In this paper, turbulent variables $k$ and $\omega$ are updated separately from the conservative variables in the GKS. 
Because turbulent modeling error is dominant in RANS simulation, there is no special high-order reconstruction for Eq.(\ref{k_w_sst}). In current study, incorporated with second-order GKS for conservative flow variables, the turbulent equations are solved
with first-order upwind reconstruction and Roe scheme \cite{roe1981approximate} for advection terms. When coupled with the high-order GKS, the turbulent models are solved by WENO-JS reconstruction and Roe scheme. Considering the source terms are quiet stiff for $k - \omega$ SST model, second-order central difference is used for source terms in Eq.(\ref{k_w_sst}).

\section{Numerical tests}

In this section, numerical tests from  low-speed smooth flow to transonic flow will be presented to validate our numerical scheme. The collision time takes
\begin{align*}
	\tau + \tau_t = \frac{\mu + \mu_t}{p} + C \frac{|p_l - p_r|}{|p_l + p_r|} \Delta t,
\end{align*}
where $\mu$ is the viscous coefficient obtained from Sutherland's Law, $\mu_t$ comes from the turbulence model, and $p$ is the pressure at the interface. $C$ is set to be $1.5$ in the computation, $p_l$ and $p_r$ denote the pressures on the left and right sides at the cell interface. $\Delta t$ is the time step which is determined according to the CFL condition.

\subsection{LES case: incompressible decaying homogeneous isotropic turbulence}
Incompressible decaying homogeneous isotropic turbulence (IDHIT) is the most fundamental turbulent flow, a classical system for turbulence theory \cite{sagaut2008homogeneous}. Additionally, IDHIT is widely used to validate turbulence model and is regarded as a benchmark to verify the performance of high-order scheme. In current study, the reference experiment is conducted by Comte-Bellot et al. \cite{comte1971simple}, with Taylor Reynolds number $Re_{\lambda} = 71.6$ and turbulent Mach number $Ma_t = 0.2$.  In numerical simulation, computation domain is a $(2\pi)^3$ box with $64^3$ and $128^3$ uniform grids, and periodic boundary condition in $6$ faces are applied. The initial turbulent fluctuating velocity fields is computed from experimental energy spectral, with constant pressure, density and temperature.

The turbulent fluctuating velocity  $u^{'}$, the Taylor microscale $\lambda$, the Taylor Reynolds number $Re_{\lambda}$, the turbulent Mach number $Ma_t$, and the spectral of turbulence kinetic energy (TKE) are defined as
\begin{align*}
	u^{'} &= <(u_1^2 + u_2^2 + u_3^2)/3>^{1/2}, \\
	\lambda^2 &= \frac{{u^{'}}^2}{<(\partial u_1/ \partial x_1)^2>}, \\
	Re_{\lambda} &= \frac{u^{'} \lambda}{\nu}, \\
	Ma_t &= \frac{<u_1^2 + u_2^2 + u_3^2>^{1/2}}{a}, \\
	E(\kappa) &= \frac{1}{2}\int_{\kappa_{min}}^{\kappa_{max}} \Phi_{ii}(\bm{\kappa}) \delta(|\bm{\kappa}| - \kappa) d \bm{\kappa},
\end{align*}
where $<\cdots>$ represents the space average in computation domain.$u_1$, $u_2$, and $u_3$ are three components for turbulent fluctuating velocity, $a$ represents the local sound speed, and $\nu$ represents the kinematic viscosity as $\mu/\rho$. Velocity spectral $\Phi_{ii}$ is the Fourier transform of two-point correlation, with wave number $\kappa_{min} = 0$ and $\kappa_{max} = N/2$, where $N$ is the number of grid points in each direction of box.
\begin{figure}[htp]
	\centering
	\includegraphics[height=0.5\textwidth]{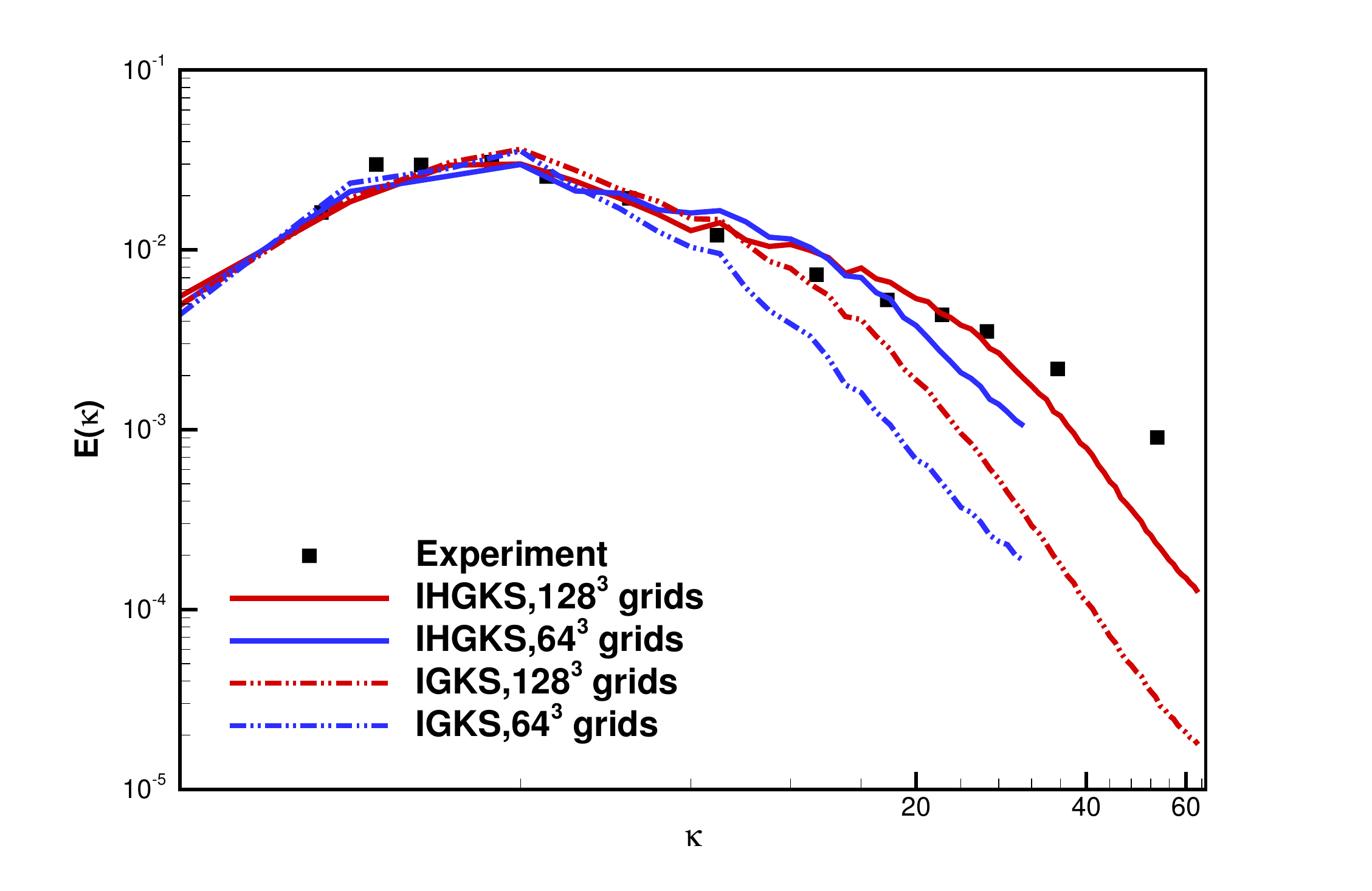}
	\caption{Spectral of TKE at dimensionless time $t^{\ast}=0.87$ with the experiment data, the IHGKS and the second-order IGKS.}
	\label{dhit_tke}
\end{figure}

For this unsteady flow, dual-step method has not been implemented in current implicit scheme, so CFL number is adopted similar as the explicit scheme $CFL = 0.15$. Current IHGKS and the second-order implicit GKS (IGKS) with Vreman-type LES model on $64^3$ and $128^3$ grids are performed, using the smooth flux as Eq.(\ref{formalsolution_smooth}). Figure \ref{dhit_tke} shows the spectral of TKE at dimensionless time $t^{\ast}=0.87$, with the IHGKS and the second-order IGKS. All schemes behave well in low wavenumber region, which means large-scale turbulent structure are resolved.  With the same grid, from the moderate wavenumber region to the high wavenumber region, TKE spectral from the high-order scheme is much closer to the experimental result, which outweighs results from the second-order scheme. In particular, it is clear that result from the high-order scheme on $64^3$ grids even performs better than that from the second-order scheme on $128^3$ grids. TKE spectral indicates that current IHGKS obtains more accurate turbulent flow fields on coarse $64^3$ grids, and second-order GKS is more dissipative.  Balancing well between the accuracy and computational costs, the IHGKS without Gaussian points is an appropriate trade-off for following engineering turbulence simulation.

\subsection{RANS 2D case: incompressible turbulence with zero pressure gradient over a flat plate}
Two-dimensional zero pressure gradient turbulence over flat plate is used to test the high efficiency of current IHGKS compared with the explicit HGKS. This is one of turbulence model verification test cases provided by the NASA turbulence modeling resource (TMR) \cite{NASA_Langley}. In current case, free stream condition is Mach number $Ma = 0.2$, and Reynolds number $Re = 5.0 \times 10^6$ with reference length $1.0$. The computational domain and boundary conditions are adopted as the NASA's website. As presented in table \ref{plate_grid}, CFL3D is implemented on fine grid G2 which provides the reference results, while the IHGKS and the explicit HGKS are performed on moderate grid G1. The total grid of G2 is almost $4$ times more than that of G1, and a smaller $Y^{+}_{plate}$ is used in G2. Here, $Y^{+}_{plate}$ is the non-dimensional wall distance for the first level grid upon the plate wall.
\begin{table}[htp]
	\caption{Grid information of moderate grid G1 and fine grid G2}
	\centering
	\begin{tabular}{c|cccc}
		\hline
		\hline
		Solver	        & Grid  & Nx $\times$ Ny   & Total grid & $Y^{+}_{plate}$  \\
		\hline
		IHGKS/HGKS      &G1  & $273 \times 193$ & $5.26 \times 10^4$      & $0.2$    \\
		\hline
		CFL3D           &G2  & $543 \times 385$ & $2.10 \times 10^5$      & $0.08$   \\
		\hline
		\hline
	\end{tabular}
	\label{plate_grid}
\end{table}

\begin{figure}[htp]
	\centering
	\subfigure[]
	{
		\includegraphics[width=0.47\textwidth]{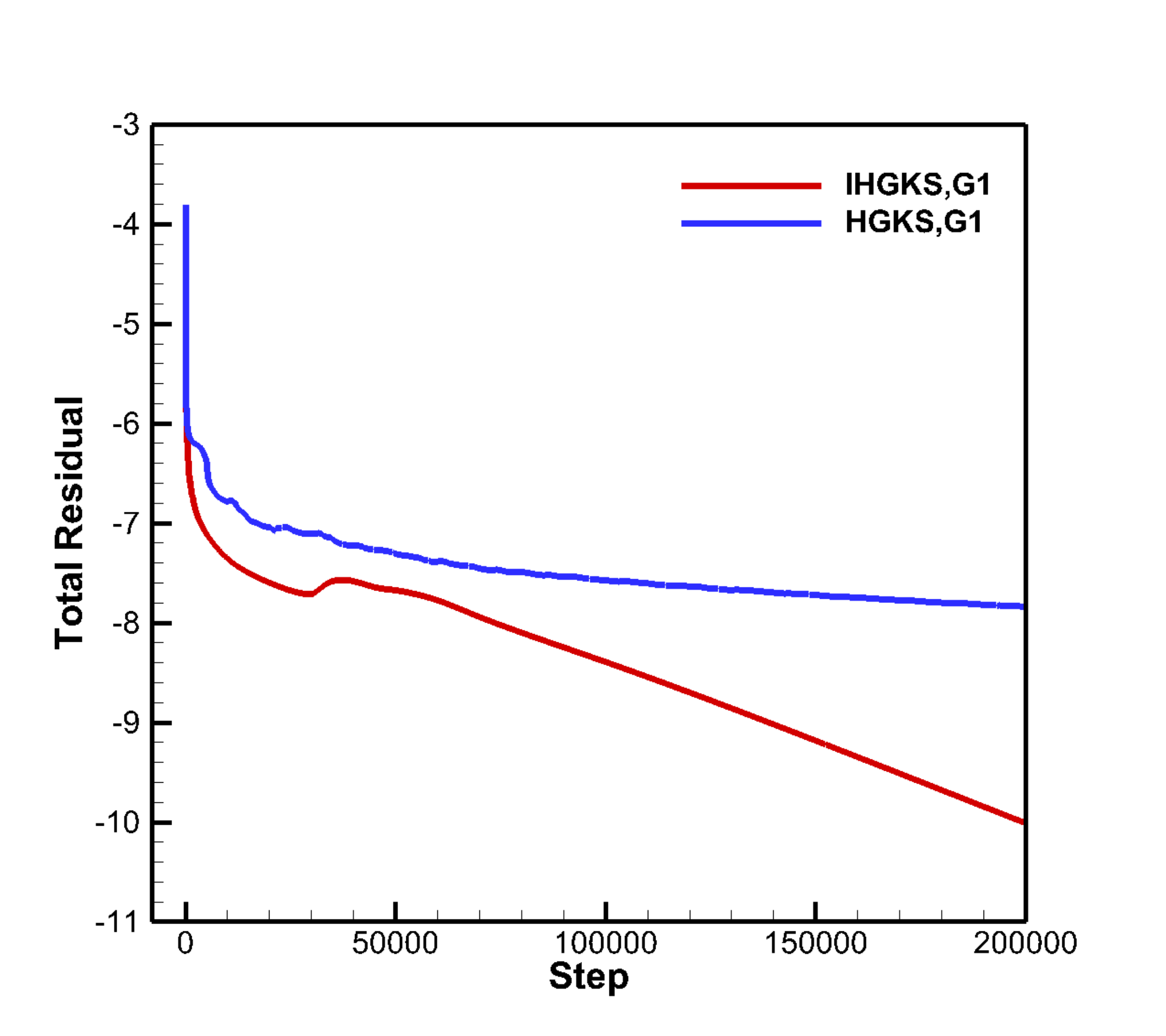}
	}
	\subfigure[]
	{
		\includegraphics[width=0.47\textwidth]{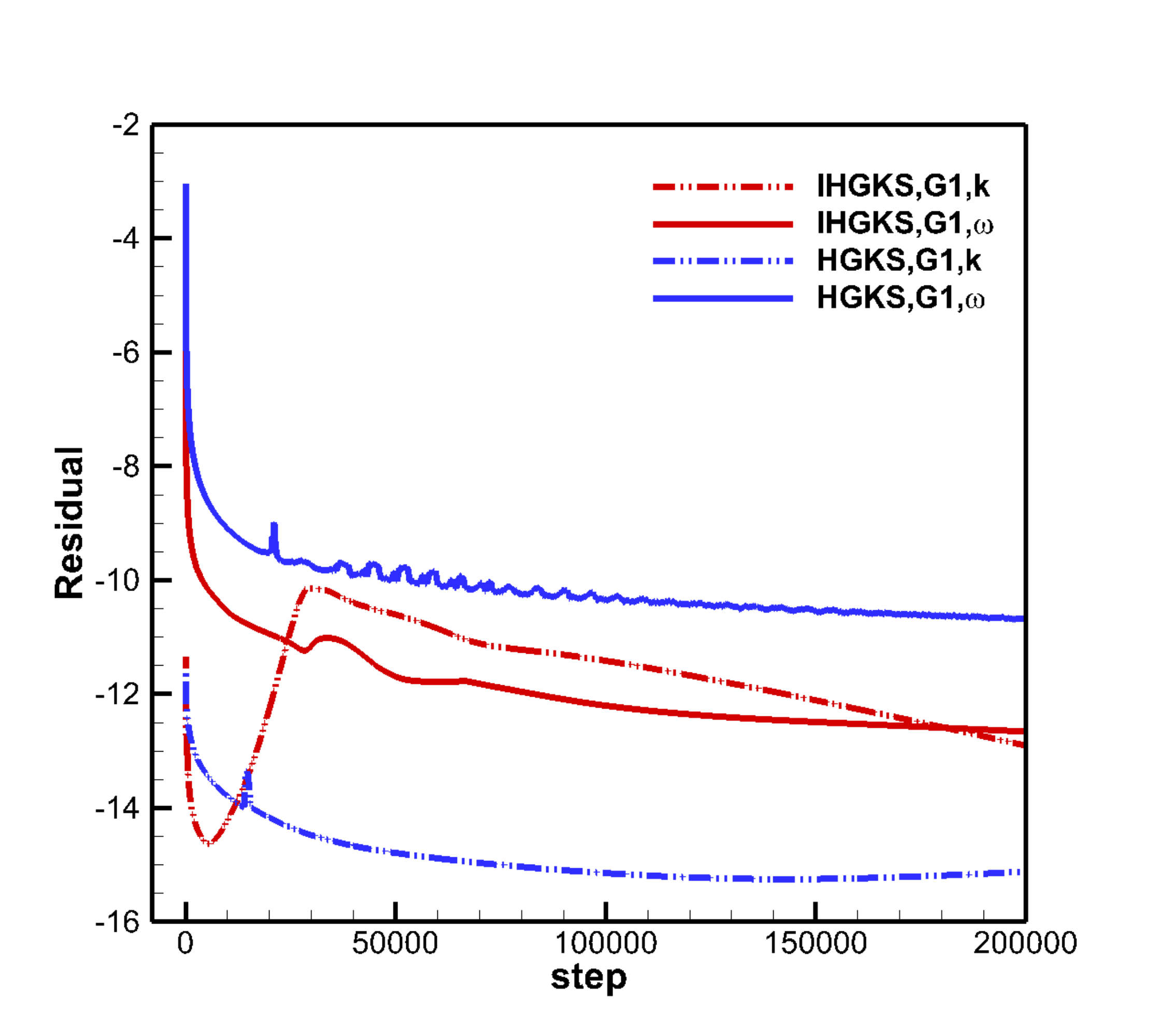}
	}
	\caption{Total residual and residual of $k - \omega$ convergence curves for the IHGKS and the explicit HGKS.}
	\label{plate_resdisual}
\end{figure}

Moderate grid G1 is split into $5$ blocks for parallel computing  on Intel Xeon E5-2962 v2 cores. As table \ref{plate_cfl} shows, the maximum CFL number which can be used for the IHGKS is $CFL = 2.5$, however, the explicit HGKS only can reach the maximum CFL number $CFL = 0.15$. With the smooth flux as Eq.(\ref{formalsolution_smooth}), the CPU time/each step of the IHGKS is $0.56$s/each step, which is slightly longer than that of the explicit HGKS. In Figure \ref{plate_resdisual}, total residual and residual of $k - \omega$ convergence curves of these two schemes are plotted. Figure \ref{plate_resdisual} (a) shows the total residual converging rate of the IHGKS is much faster than that of the explicit HGKS. While, turbulent variables $ k - \omega$ are quiet stiff as Figure \ref{plate_resdisual} (b) presents. Taking the CPU time/each step and the total residual converging rate into consideration, the IHGKS can speed up more than $10$ times than the explicit HGKS. The significant acceleration on computational efficiency obtained by implicit LU-SGS method is pretty important when implementing engineering turbulence using high-order GKS. In the following cases, considering the affordable computational costs, only the IHGKS and the second-order IGKS will be implemented for high-Reynolds number turbulent flows.
\begin{table}[htp]
	\caption{Maximum CFL number for the IHGKS and the explicit HGKS}
	\centering
	\begin{tabular}{c|ccc}
		\hline
		\hline
		Solver     &Grid          & CFL number      &CPU time\\
		\hline
		IHGKS      &G1            & 2.50            & $0.56$s/each step\\
		\hline
		HGKS       &G1            & 0.15            & $0.51$s/each step\\
		\hline
		\hline
	\end{tabular}
	\label{plate_cfl}
\end{table}

\begin{figure}[htp]
	\centering
	\subfigure[]
	{
		\includegraphics[width=0.47\textwidth]{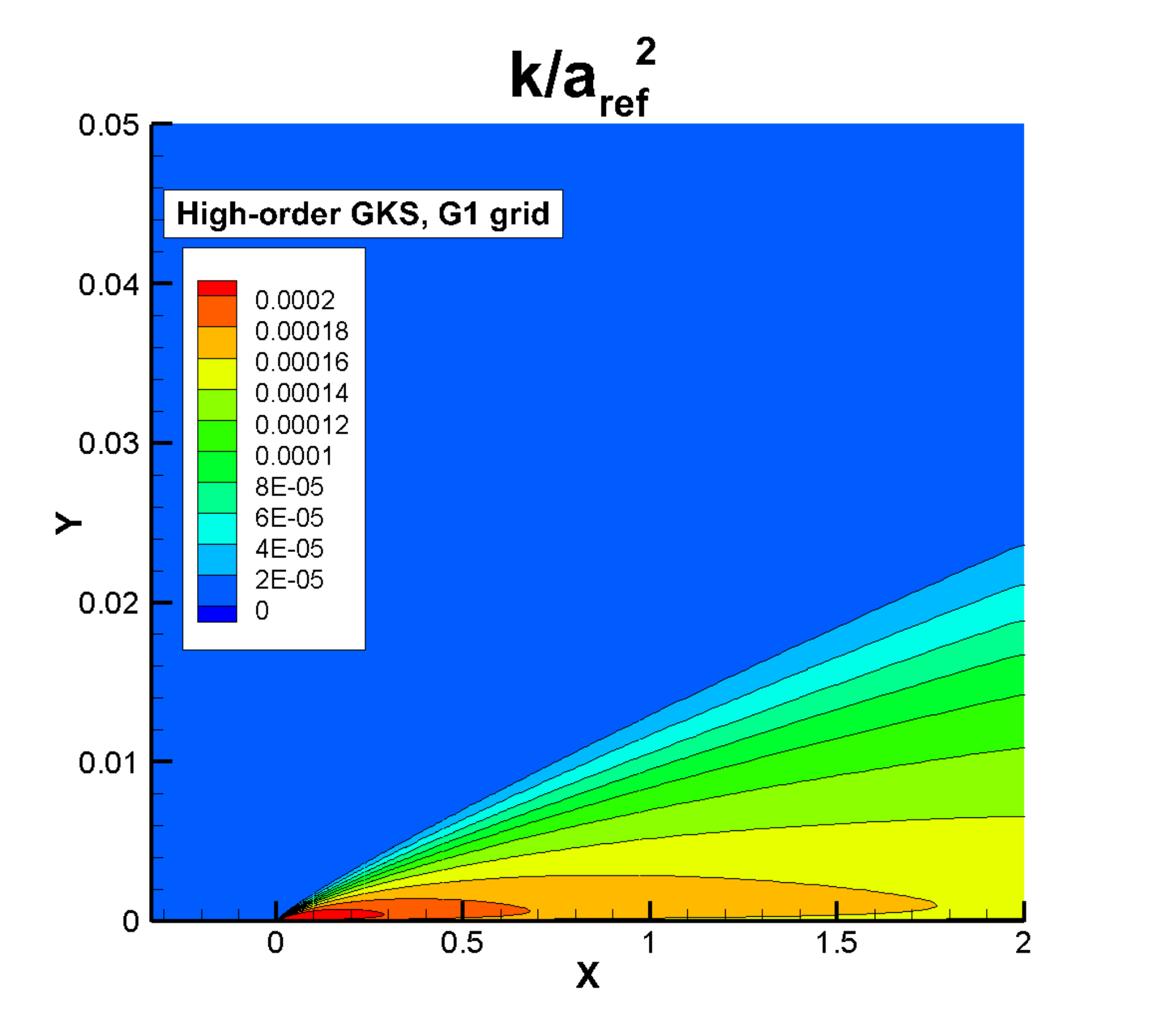}
	}
	\subfigure[]
	{
		\includegraphics[width=0.472\textwidth]{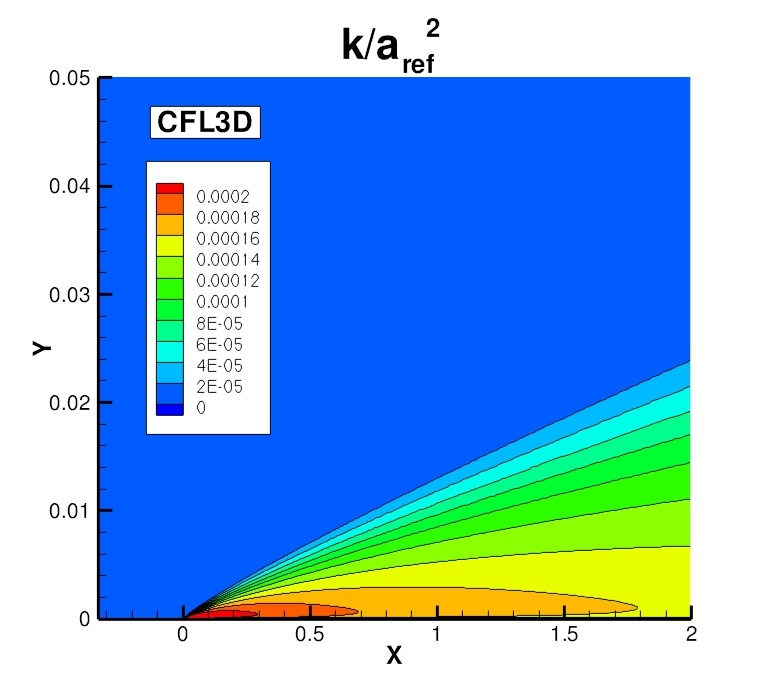}
	}	
	\subfigure[]
	{
		\includegraphics[width=0.47\textwidth]{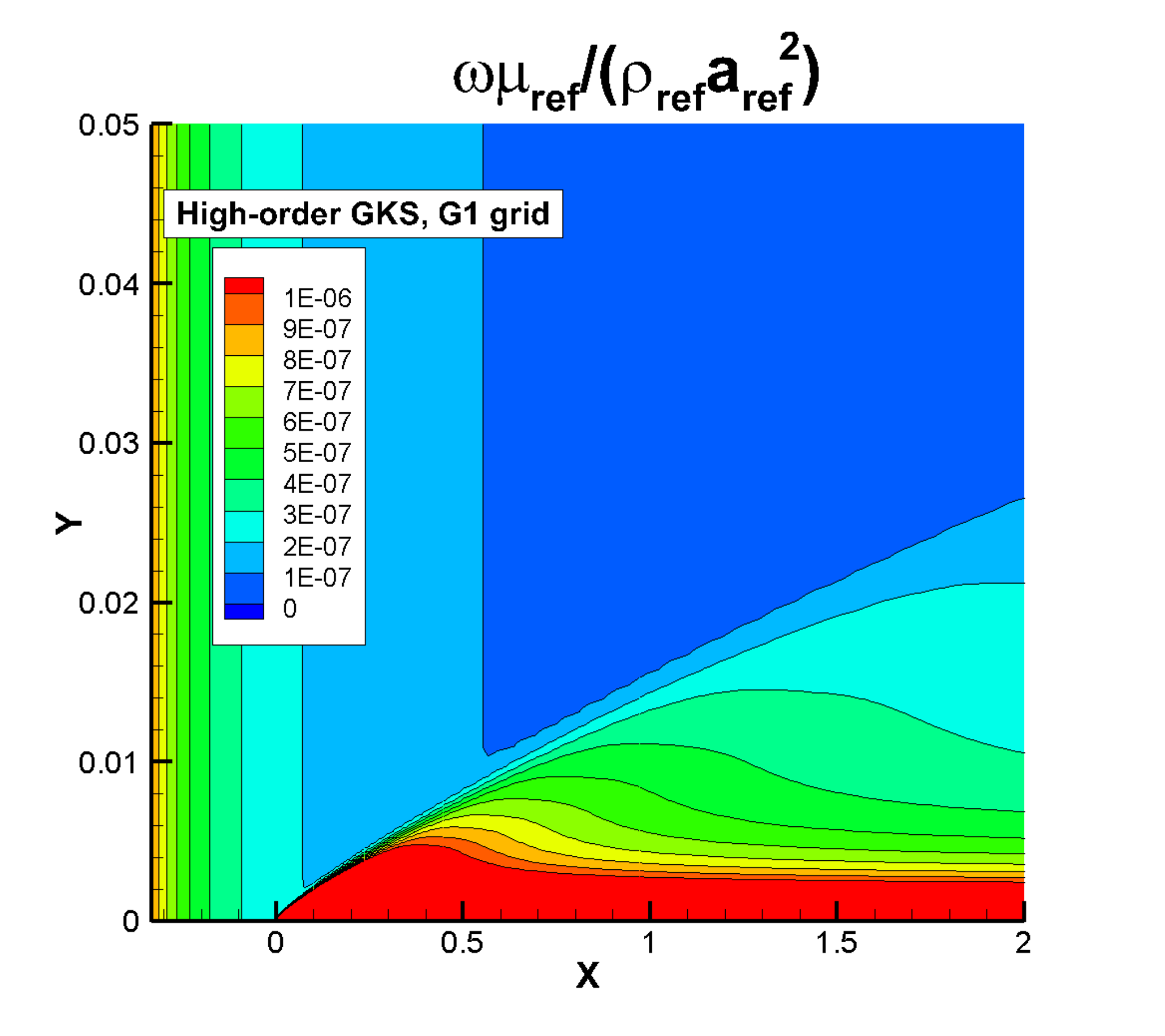}
	}
	\subfigure[]
	{
		\includegraphics[width=0.472\textwidth]{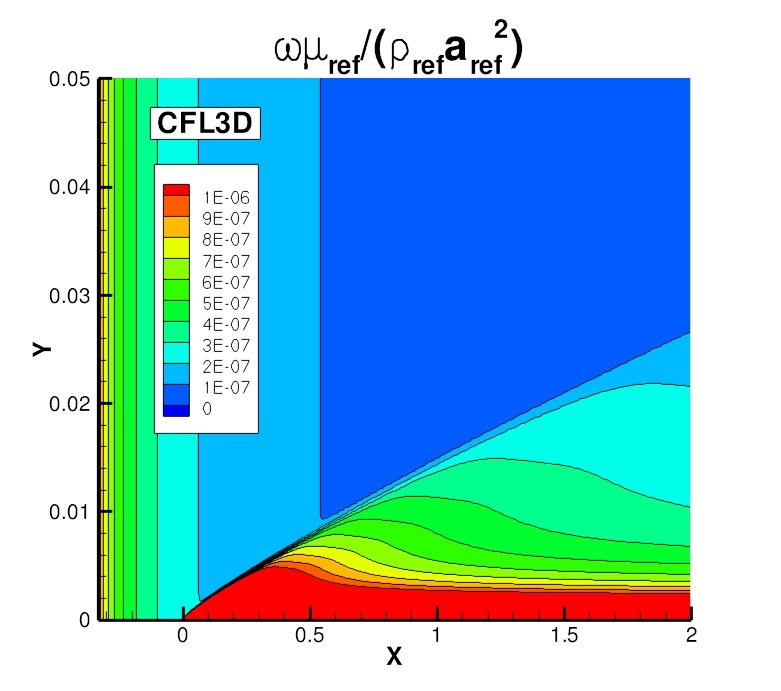}
	}
	\caption{Non-dimensional contours of TKE $k$ and specific dissipation rate $\omega$, (a)(c) from the IHGKS on moderate grid G1, and (b)(d) from the CFL3D on fine grid G2.}
	\label{plate_k_omega}
\end{figure}

\begin{figure}[htp]
	\centering
	\subfigure[]
	{
		\includegraphics[width=0.47\textwidth]{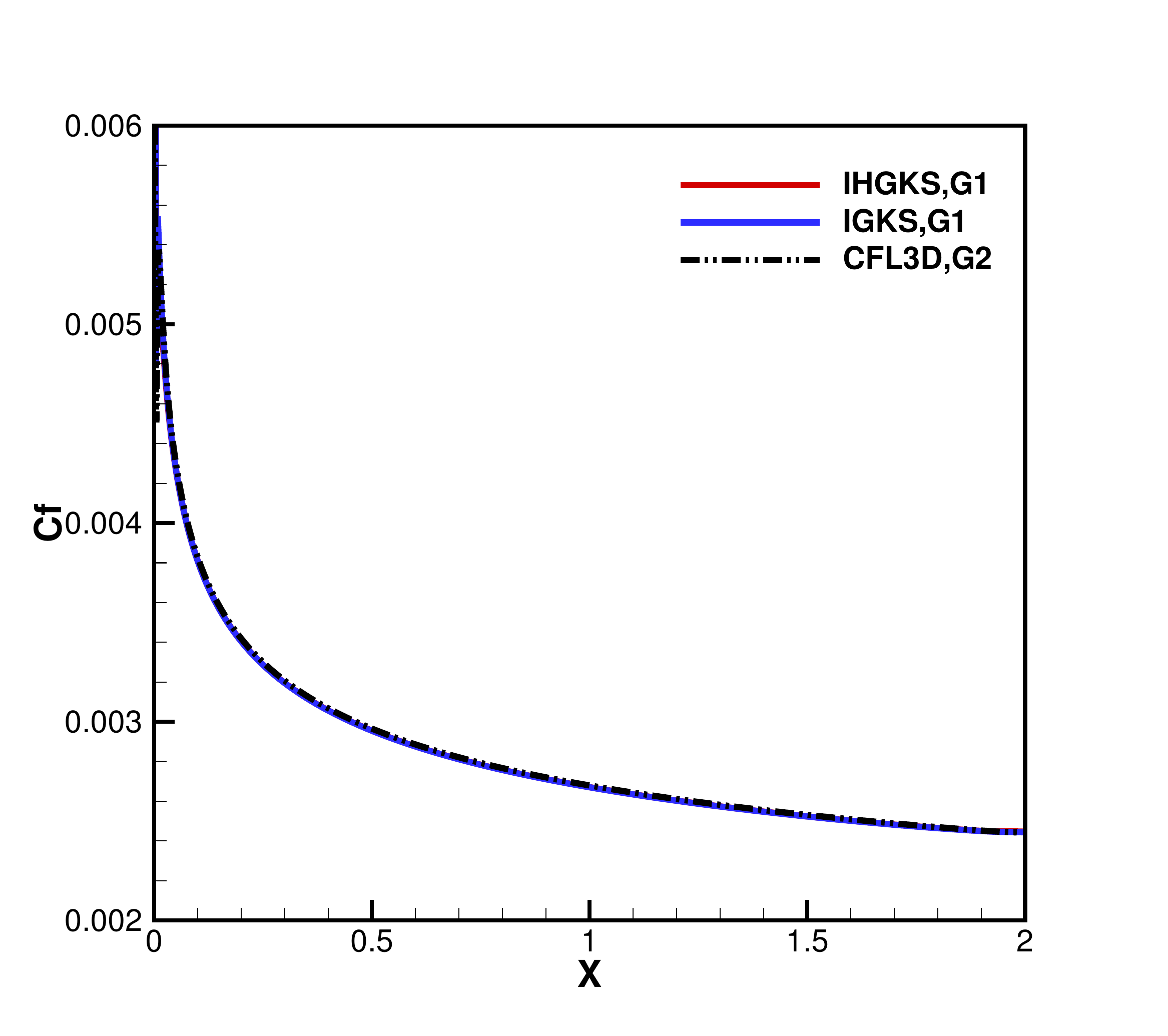}
	}
	\subfigure[]
	{
		\includegraphics[width=0.47\textwidth]{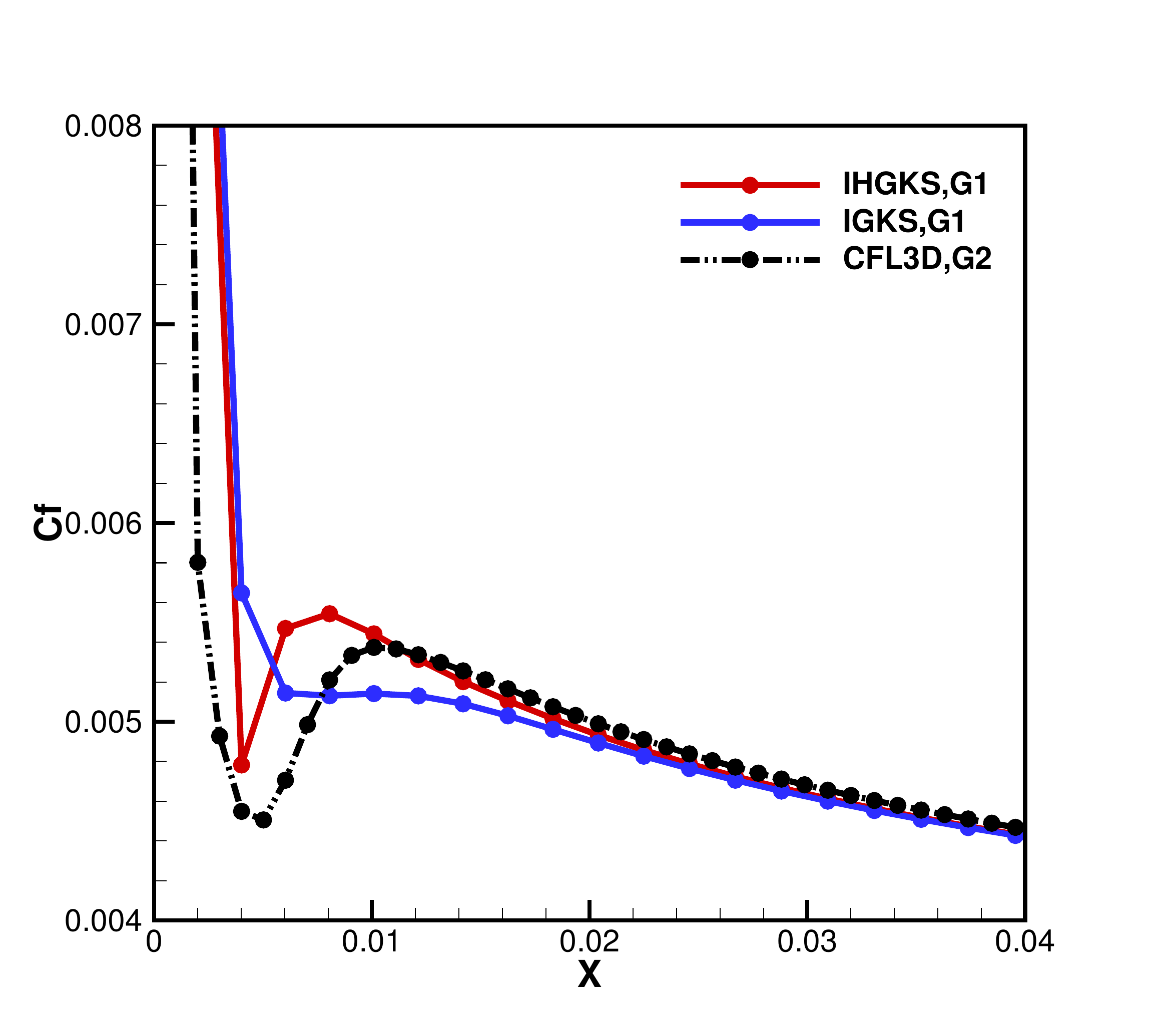}
	}
	\caption{Skin friction coefficient $C_f$ along the flat plate (a) and within the local transitional region (b) from the current IHGKS, the second-order IGKS, and the second-order CFL3D.}
	\label{plate_cf}
\end{figure}

To validate current implementation of $k - \omega$ SST model, it is legitimate to compare the TKE $k$ and specific dissipation rate $\omega$ with reference solutions firstly. Figure \ref{plate_k_omega} shows that the non-dimensional TKE $k$ and specific dissipation rate $\omega$ of the IHGKS on moderate grid G1 agree well with the CFL3D on fine grid G2. Friction coefficient is provided for quantitative comparisons in Figure \ref{plate_cf}. Overall, skin friction coefficients along the flat plate with the IHGKS and second-order IGKS  are comparable in Figure \ref{plate_cf} (a). As shown in Figure \ref{plate_cf} (b), compared with the reference solution on fine grid G2, current IHGKS predicts the transition region well, roughly from leading edge $X = 0$ to $X = 0.02$. For the second-order IGKS, this transition region has not been captured, which behaves similarly with previous second-order GKS simulation results \cite{li2016gas}. Current case not only validates the high-efficiency of the IHGKS, but also indicates that the high-accuracy flow fields obtained by the IHGKS is required on moderate grid, such as transition prediction.

\subsection{RANS 2D case: incompressible turbulence around NACA0012 airfoil}
\begin{figure}[!htp]
	\centering
	\subfigure[]
	{
		\includegraphics[width=0.47\textwidth]{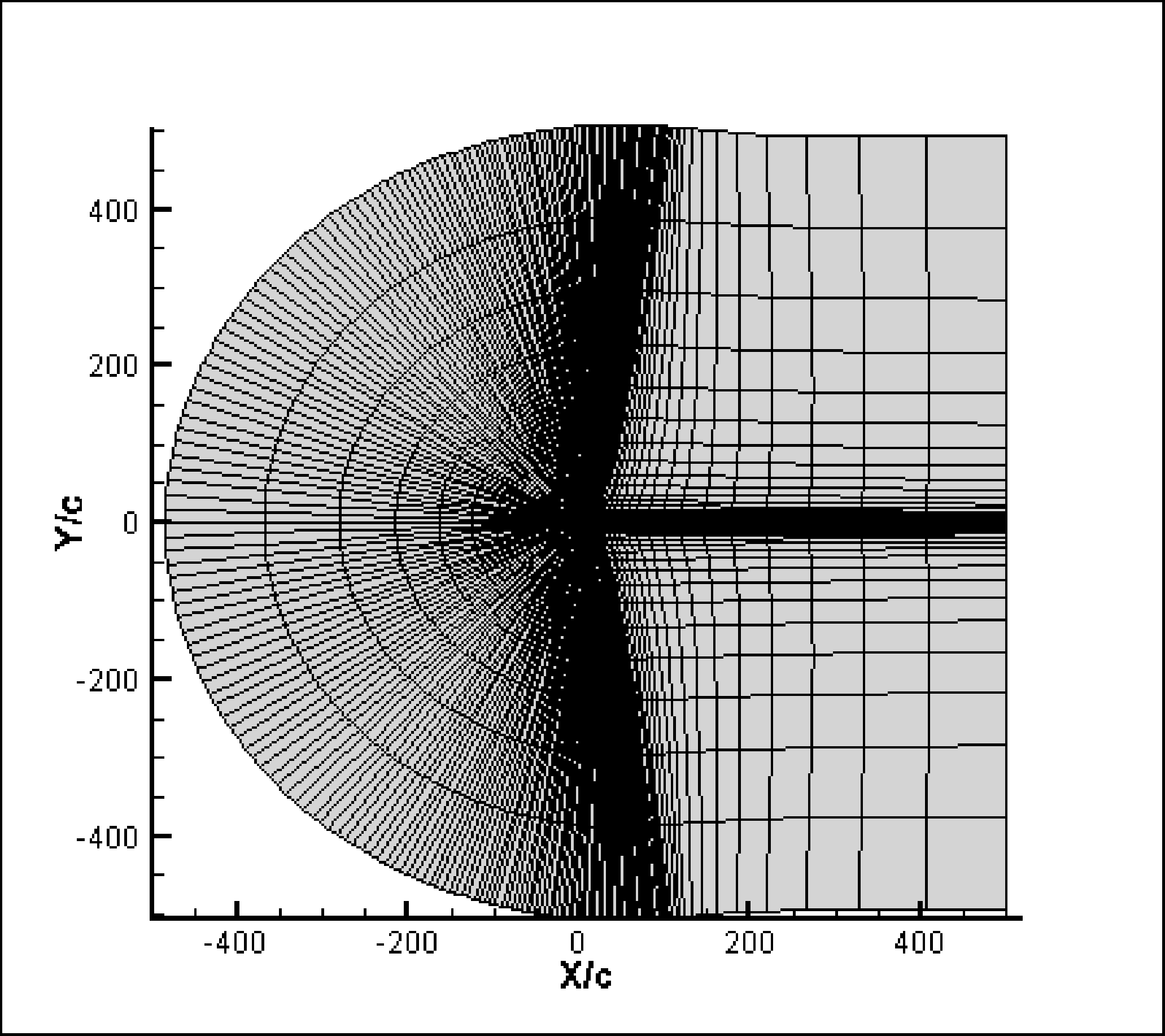}
	}
	\subfigure[]
	{
		\includegraphics[width=0.47\textwidth]{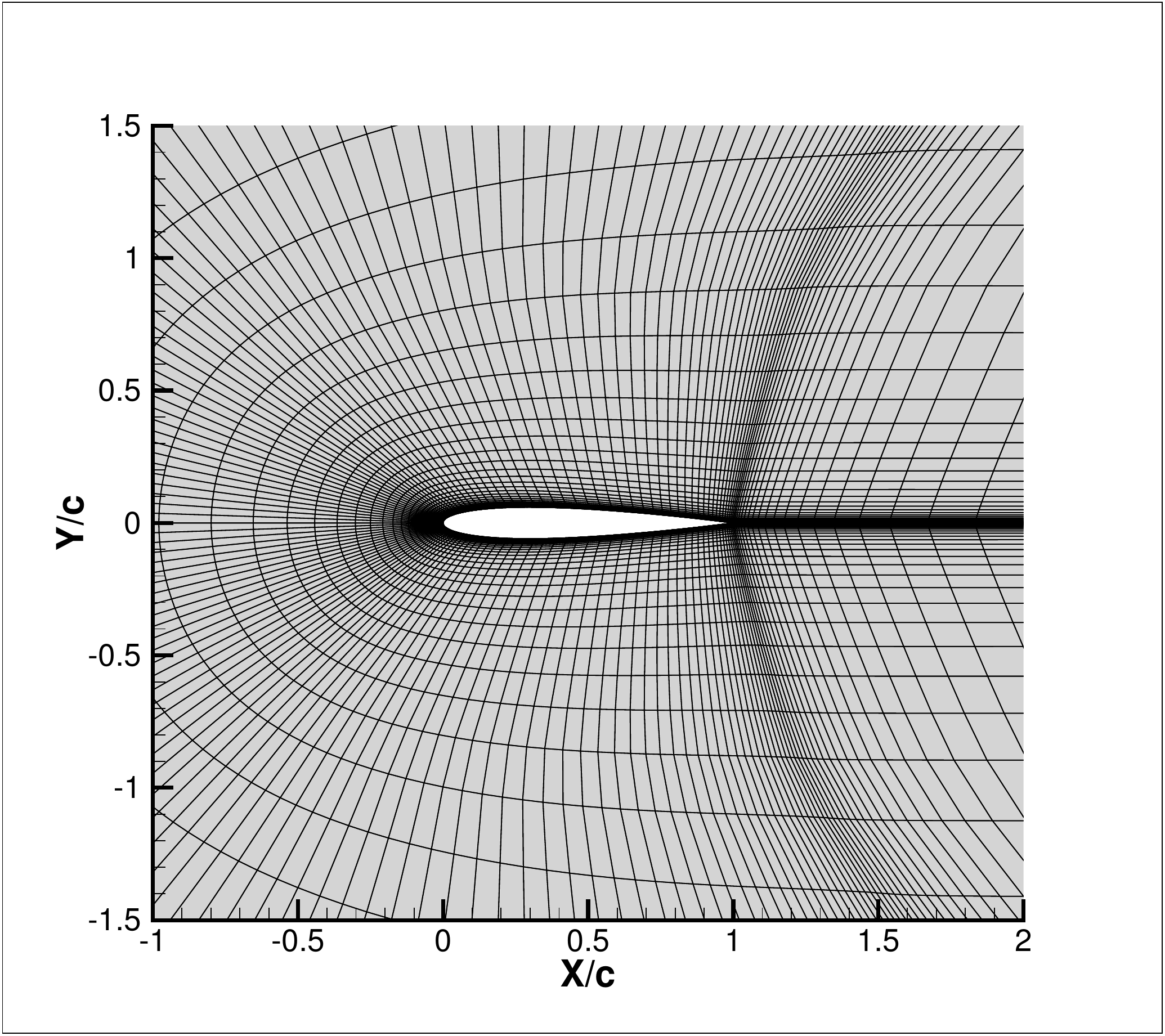}
	}
	\caption{\label{NACA0012_grid_gl}Global arrangement (a) and local arrangement (b) of coarse grid G3 around NACA0012 airfoil. $c$ is the chord length.}
\end{figure}

Engineering simulation of the incompressible NACA0012 airfoil with angle of attack $\alpha = 15^o$ is implemented, and the free stream condition is Mach number $Ma = 0.15$, Reynolds number $Re = 3.0 \times 10^6$ with the reference chord length $c = 1.0$. For lift coefficient and drag coefficient study, the reference area is set as $area = 1$. This is another turbulence model verification test case provided by the NASA TMR \cite{NASA_Langley}, which involves wall curvature, and thus pressure gradients are no longer equal to zero as above incompressible flat plate turbulence. The computational domain and boundary conditions are used as the the NASA's website. As table \ref{NACA0012_grid}, current study is based on the coarse grid G3 with IHGKS/IGKS, and the referee data are from the fine grid G4 with CFL3D. The total grid of G4 is almost $16$ times more than that of G3, and an approximate $2$ times smaller $Y^{+}_{wall}$ is used in fine grid G4. Here, $Y^{+}_{wall}$ is the non-dimensional wall distance for the first level grid upon the NACA0012 airfoil. Grid arrangement around NACA0012 is presented in Figure \ref{NACA0012_grid_gl}.
\begin{table}[htp]
	\caption{Grid information of coarse grid G3 and fine grid G4}
	\centering
	\begin{tabular}{c|ccc}
		\hline
		\hline
		Solver      &Grid      & Nx $\times$ Ny   & Total grid   \\
		\hline
		IHGKS/IGKS  &G3        & $225 \times 65$  & $1.46 \times 10^4$        \\
		\hline
		CFL3D       &G4        & $897 \times 257$ & $2.30 \times 10^5$        \\
		\hline
		\hline
	\end{tabular}
	\label{NACA0012_grid}
\end{table}

\begin{figure}[!htp]
	\centering
	\subfigure[]
	{
		\includegraphics[width=0.5\textwidth]{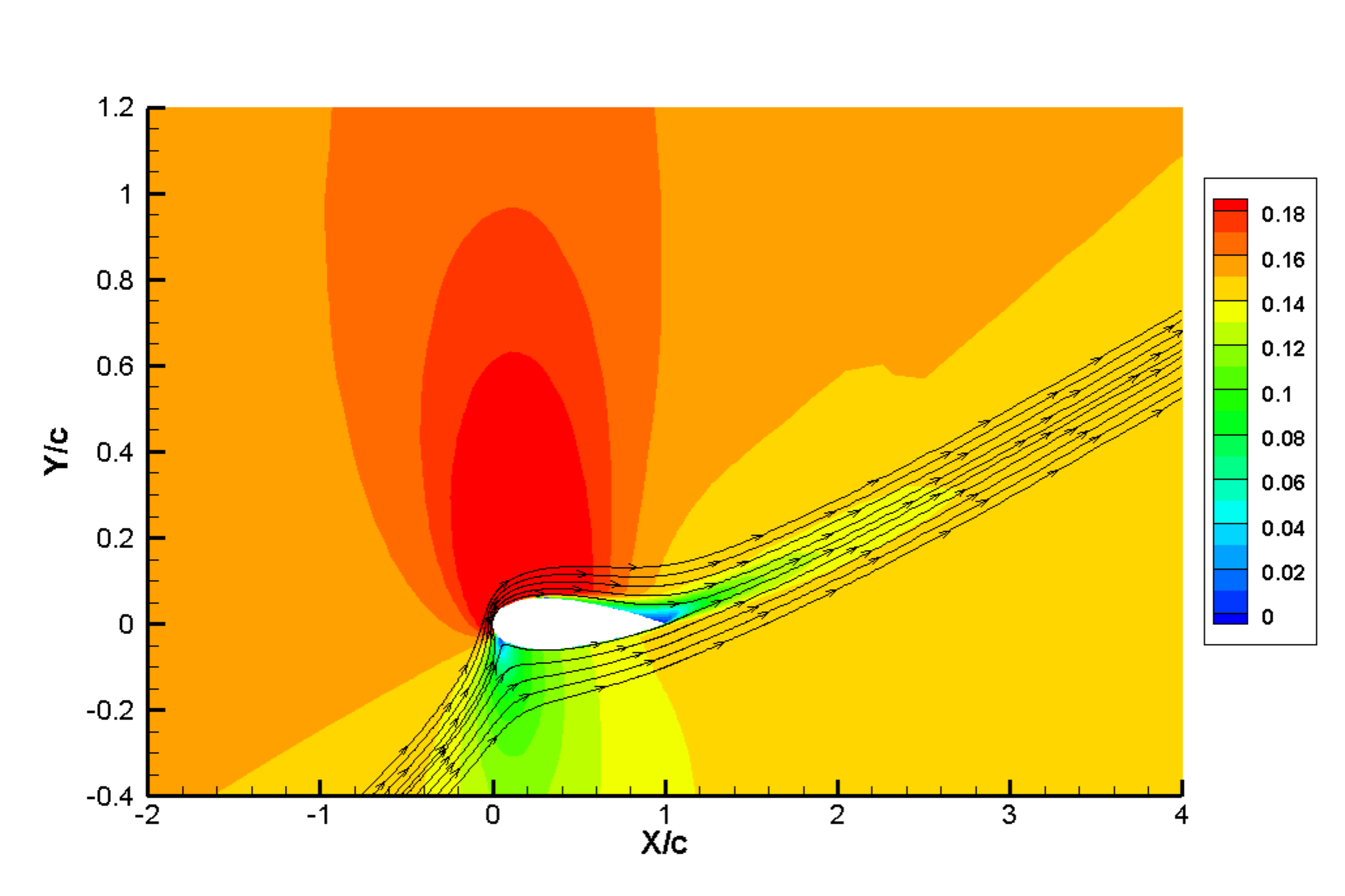}
	}
	\subfigure[]
	{
		\includegraphics[width=0.6\textwidth]{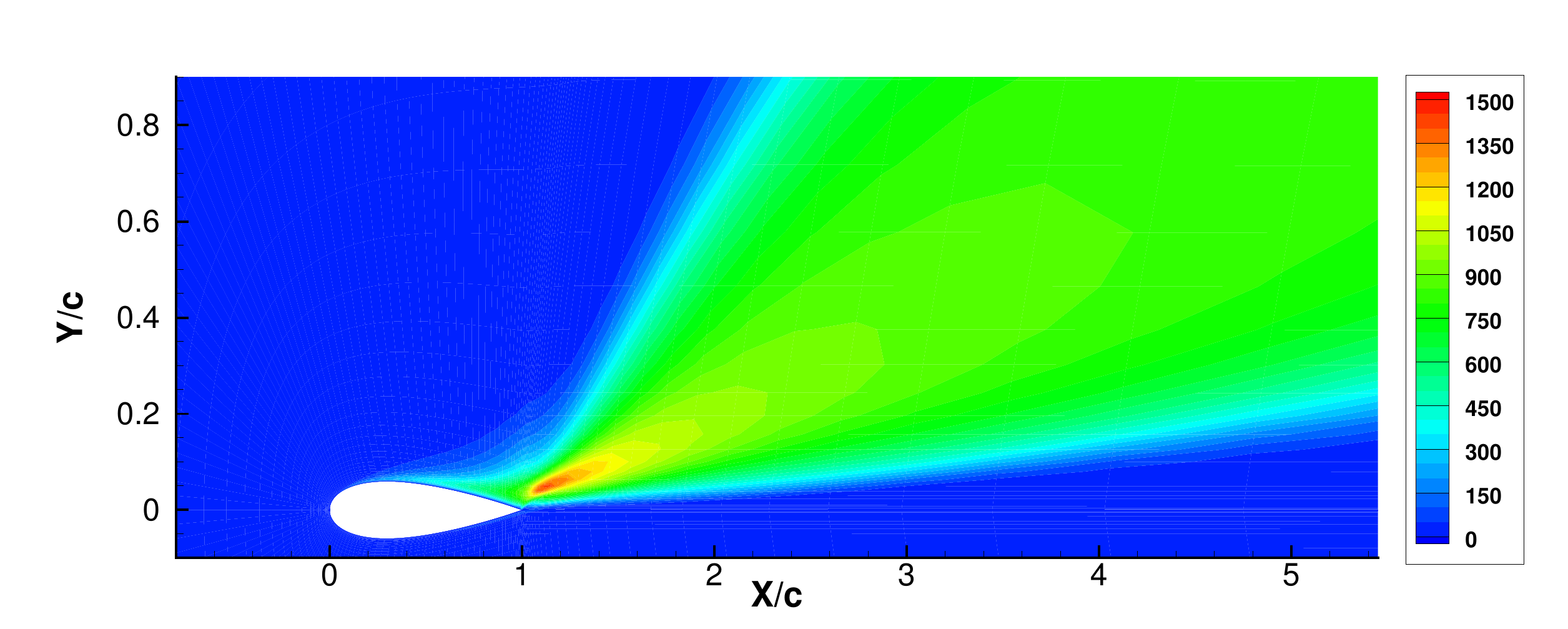}
	}
	\caption{\label{NACA0012_r3_cp_cf15} Contours of Mach number (a) and normalized viscosity $\mu_t/\mu$ (b) with the IHGKS on coarse grid G3 around NACA0012 airfoil.}
\end{figure}

\begin{figure}[!htp]
	\centering
	\subfigure[]
	{
		\includegraphics[width=0.47\textwidth]{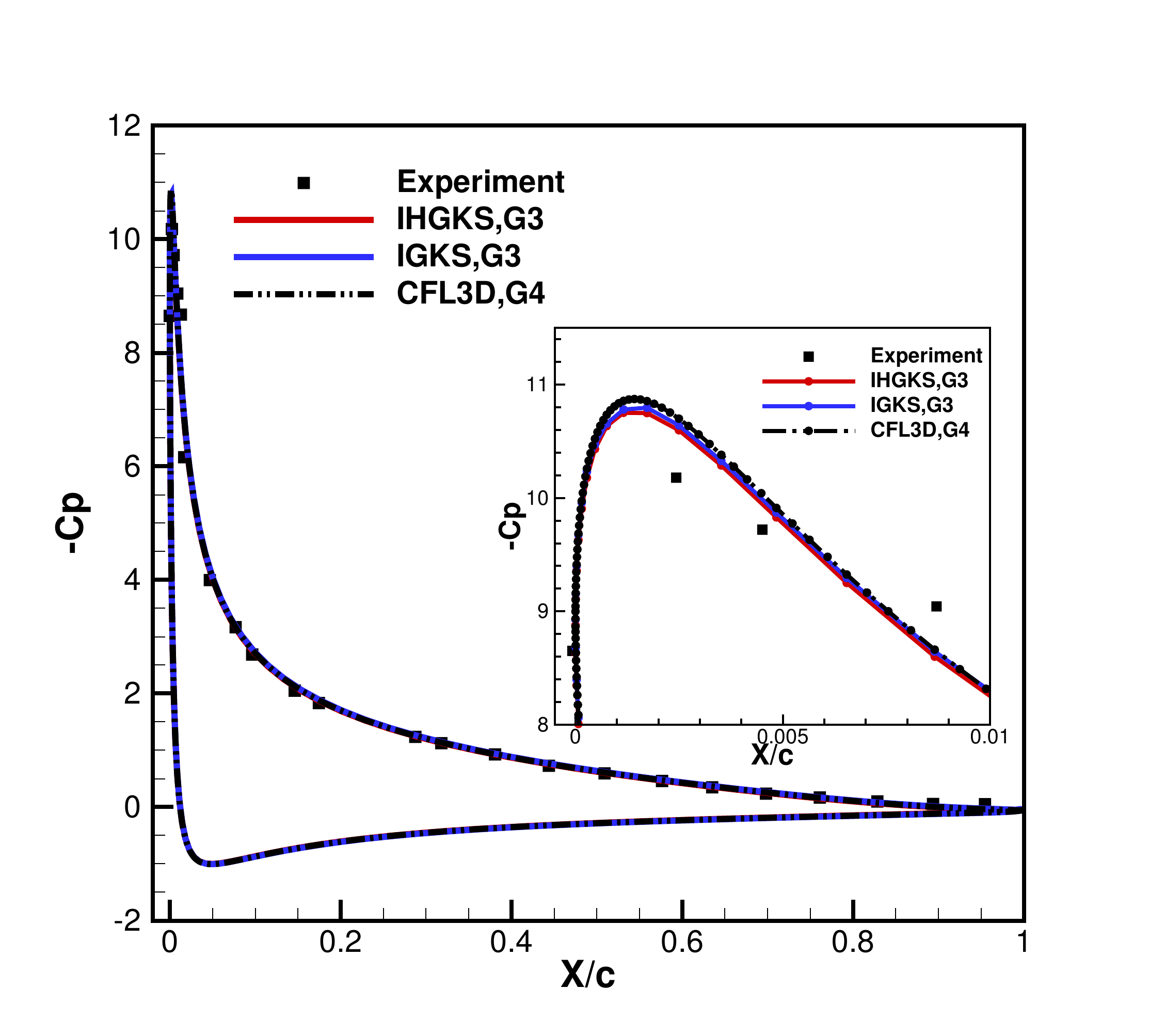}
	}
	\subfigure[]
	{
		\includegraphics[width=0.47\textwidth]{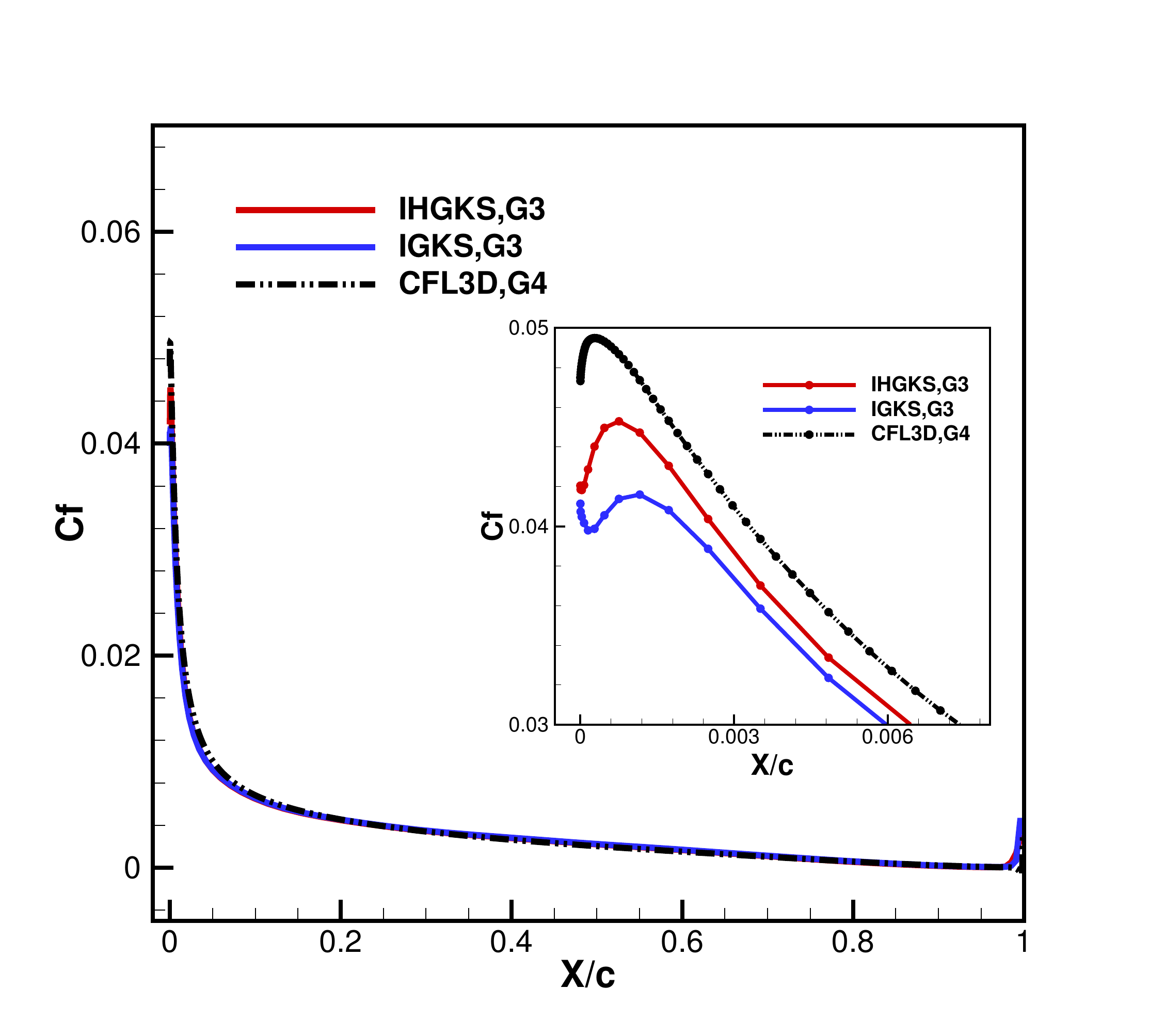}
	}
	\caption{\label{NACA0012_r3_cp_cf15_profile} Comparisons on pressure coefficient distributions $Cp$ around airfoil (a) and skin friction coefficient $C_f$ distributions on upper airfoil (b) from the experiment data, the current IHGKS, the second-order IGKS, and the second-order CFL3D.}
\end{figure}

CFL number is adopted as $CFL = 8$ and the total residual reduce down to the $6$ orders of magnitude. In Figure \ref{NACA0012_r3_cp_cf15}, Mach number contour and contour of normalized viscosity $\mu_t/\mu$ are presented  with the IHGKS on corase grid G3. As Figure \ref{NACA0012_r3_cp_cf15} shows, the incompressible flow fields are quite smooth, which is consistent with using the smooth flux as Eq.(\ref{formalsolution_smooth}). Pressure coefficient distributions around airfoil and skin friction coefficient distributions on upper airfoil are presented in Figure \ref{NACA0012_r3_cp_cf15_profile}. Among several sets of experimental pressure data provided by NASA's website, data from Gregory et al. \cite{gregory1973low} is chosen for validation in current study. It is believed that the Gregory data are likely more two-dimensional and hence more appropriate for CFD validation of surface pressures. Figure \ref{NACA0012_r3_cp_cf15_profile} (a) shows that pressure coefficients based on current IHGKS are much closer with the experimental data, which is better than the results from  the second-order IGKS. As presented in Figure \ref{NACA0012_r3_cp_cf15_profile} (b), the skin friction coefficients with the IHGKS agree quiet well with the reference data than those from the second-order IGKS. The drag coefficient $C_D$ is very sensitive to the pressure and skin friction distribution around the airfoil. As presented in Table \ref{NACA0012_CLCD}, the lift coefficient, the drag coefficient and lift-drag ratio $L/D$ from the IHGKS are very close to the reference data with CFL3D on fine grid G4. Especially, the discrepancy on $C_D$ is within $3$ drag counts (0.0001), reaching the high-level requirement for engineering turbulence simulation. For the second-order IGKS on coarse gird G3, the lift coefficient is acceptable, while it over predicts the drag coefficient $C_D$ by $17.7\%$, almost 40 drag counts. This significant improvement on drag coefficient with coarse grid, confirms the high-accuracy turbulent flow fields are obtained by current IHGKS.
\begin{table}[!htp]
	\caption{$C_L$ and $C_D$ for NACA0012}
	\centering
	\begin{tabular}{c|ccc}
		\hline \hline
		Solver              &IHGKS      &IGKS     &CFL3D\\
		\hline
		$C_L$              &1.4930     &1.5170   &1.5060 \\
		\hline
		$C_D$              &0.02198    &0.02618  &0.02224 \\
		\hline
		$L/D$              &67.93      &57.94    &67.23 \\
		\hline \hline
	\end{tabular}
	\label{NACA0012_CLCD}
\end{table}

\subsection{RANS 2D case: transonic turbulence around RAE2822 airfoil}
\begin{figure}[!htp]
	\centering
	\includegraphics[width=0.47\textwidth]{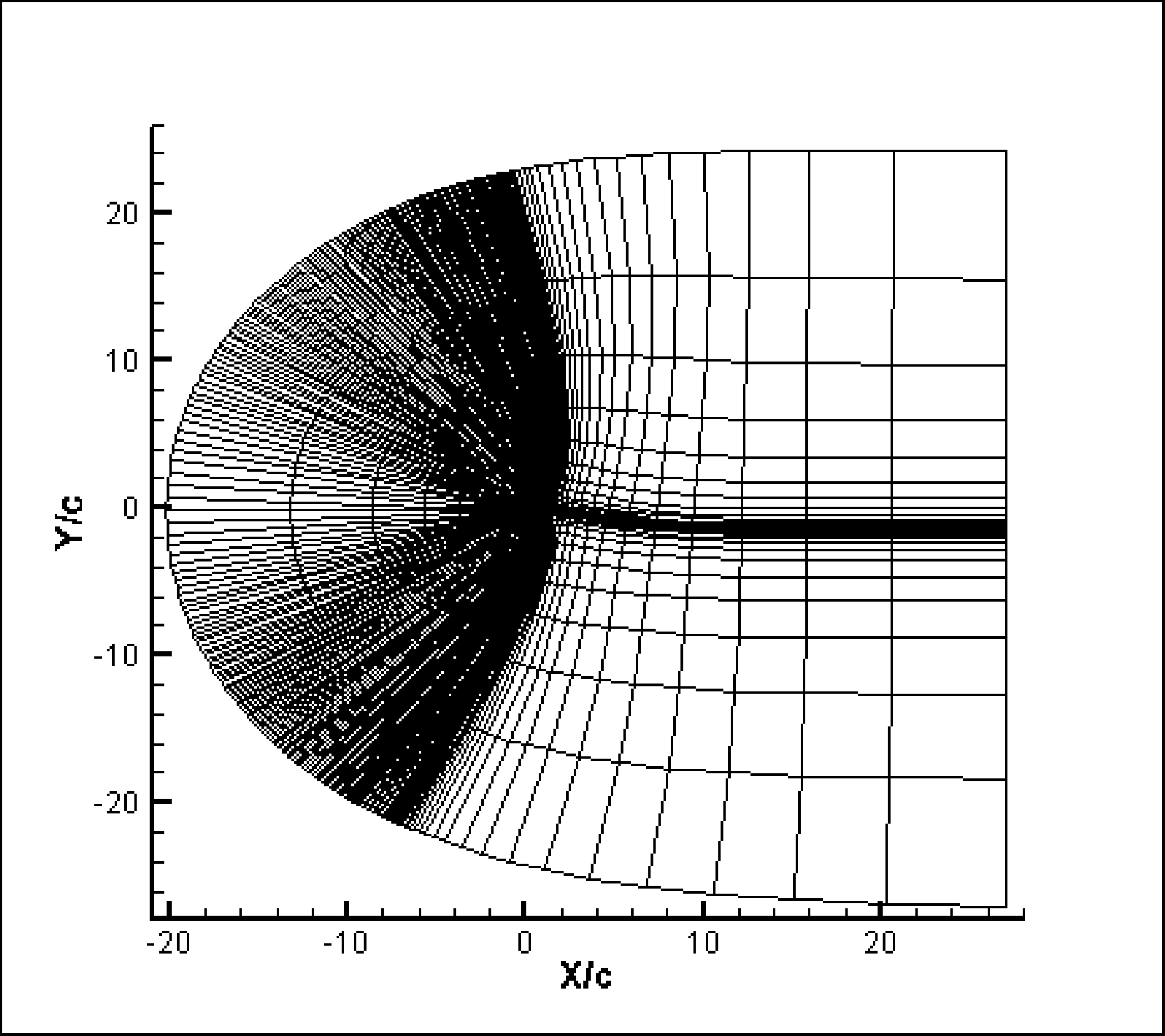}
	\includegraphics[width=0.47\textwidth]{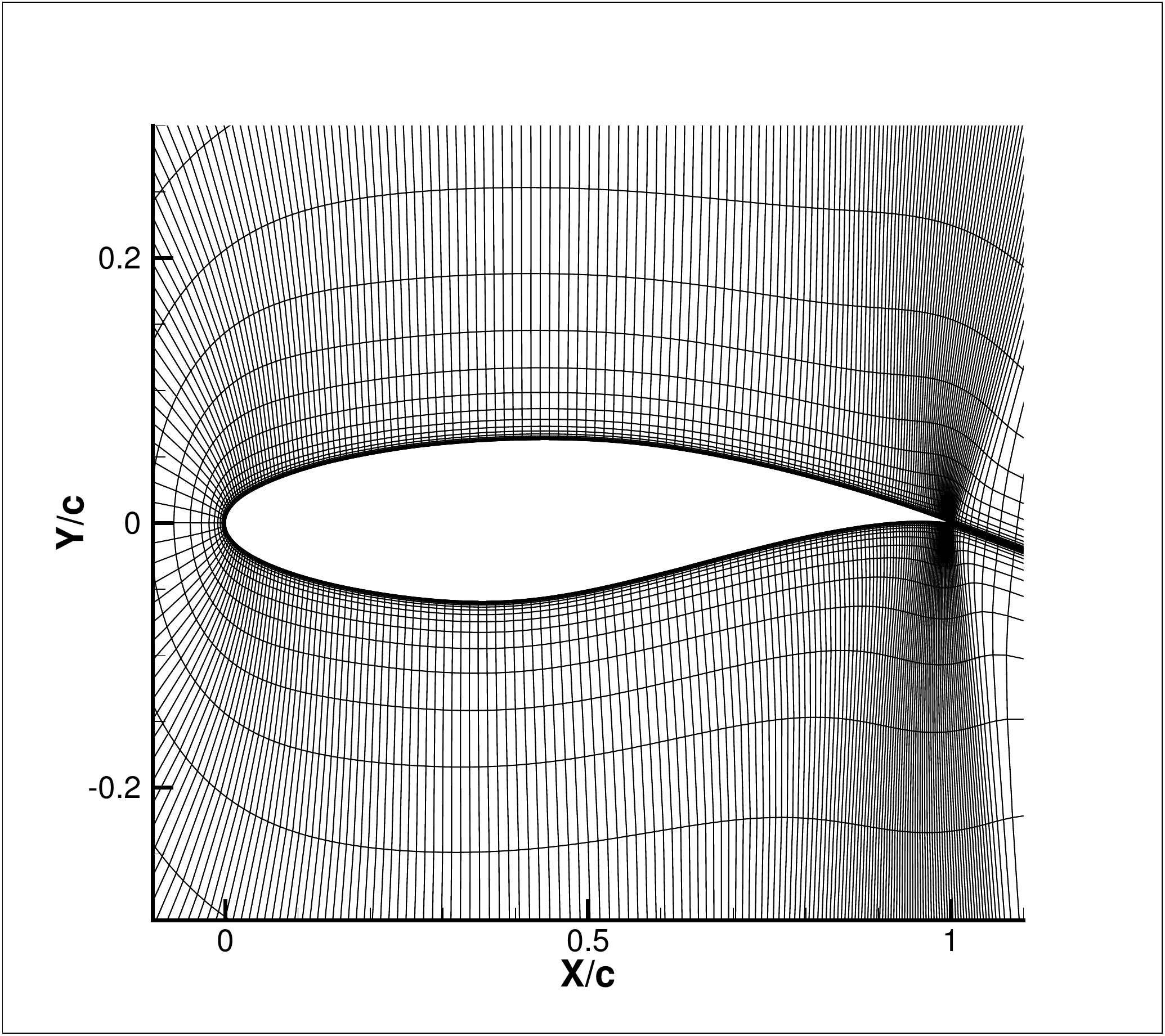}
	\caption{\label{RAE2822_grid}Global gird arrangement (a) and local grid arrangement (b) around RAE2822 airfoil.}
\end{figure}

Transonic turbulence around RAE2822 airfoil with angle of attack $\alpha = 2.79^o$ is implemented, to validate the robustness of capturing shock in transonic high-Reynolds number turbulence by the current IHGKS. The free stream condition is Mach number $Ma = 0.729$, Reynolds number $Re = 6.5 \times 10^6$ with the reference chord length $c = 1.0$. This is one turbulence model verification benchmark provided by the NPARC Alliance CFD Verification and Validation Web Site (NPARC) \cite{NPARC}. The computational domain and boundary conditions are used as the NPARC website. With $Y^{+}_{wall} = 2.5$, current validation is based on the structured grid $369 \times 65$. The global and local grid arrangement around RAE2822 are shown in Figure \ref{RAE2822_grid}.

\begin{figure}[!htp]
	\centering
	\subfigure[]
	{
		\includegraphics[width=0.47\textwidth]{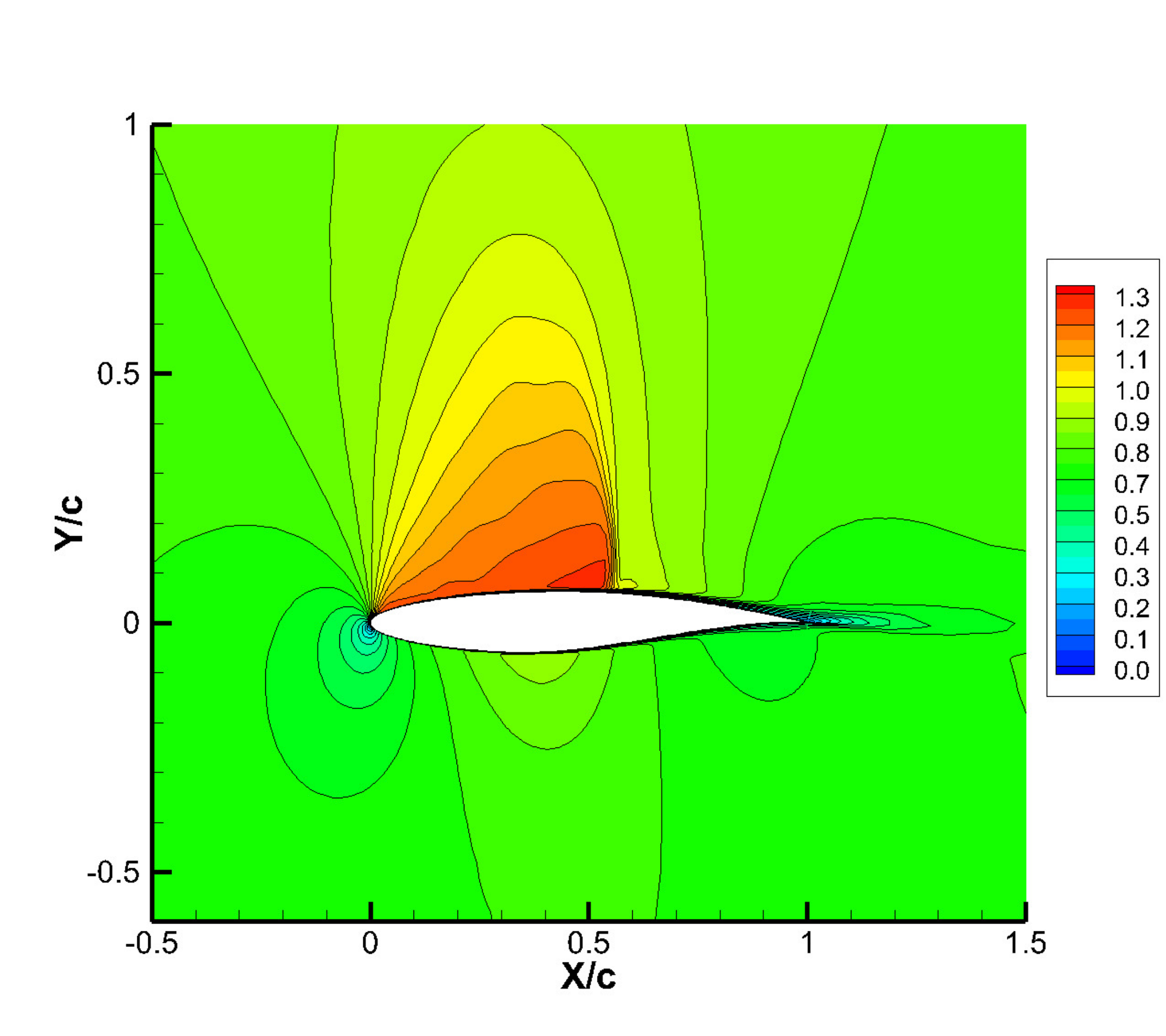}
	}
	\subfigure[]
	{
		\includegraphics[width=0.47\textwidth]{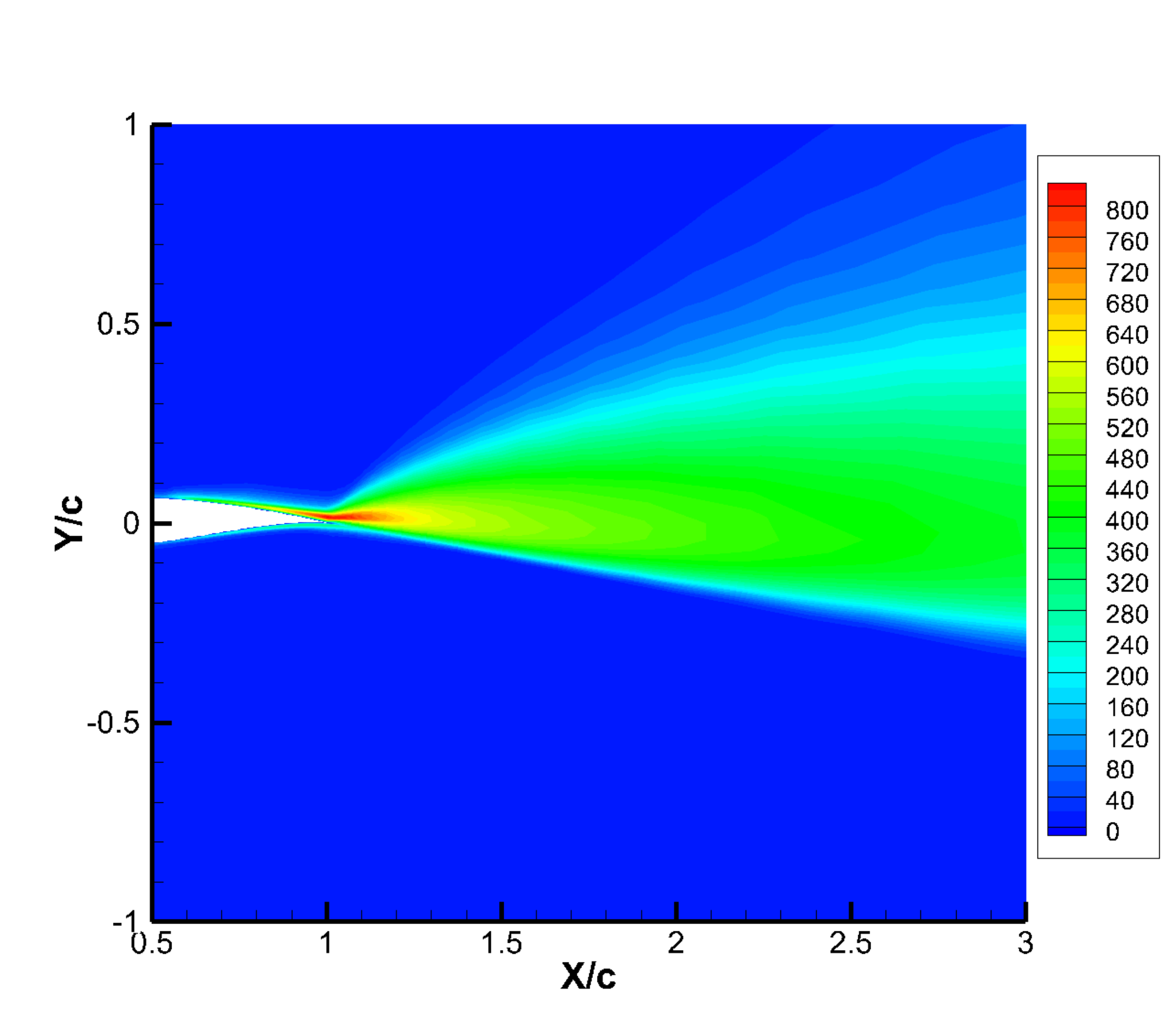}
	}
	\caption{\label{RAE2822_ma_miut} Contours of Mach number (a) and normalized viscosity $\mu_t/\mu$ (b) from the IHGKS around RAE2822 airfoil. $c$ is the chord length.}
\end{figure}

\begin{figure}[!htp]
	\centering
	\subfigure[]
	{
		\includegraphics[width=0.47\textwidth]{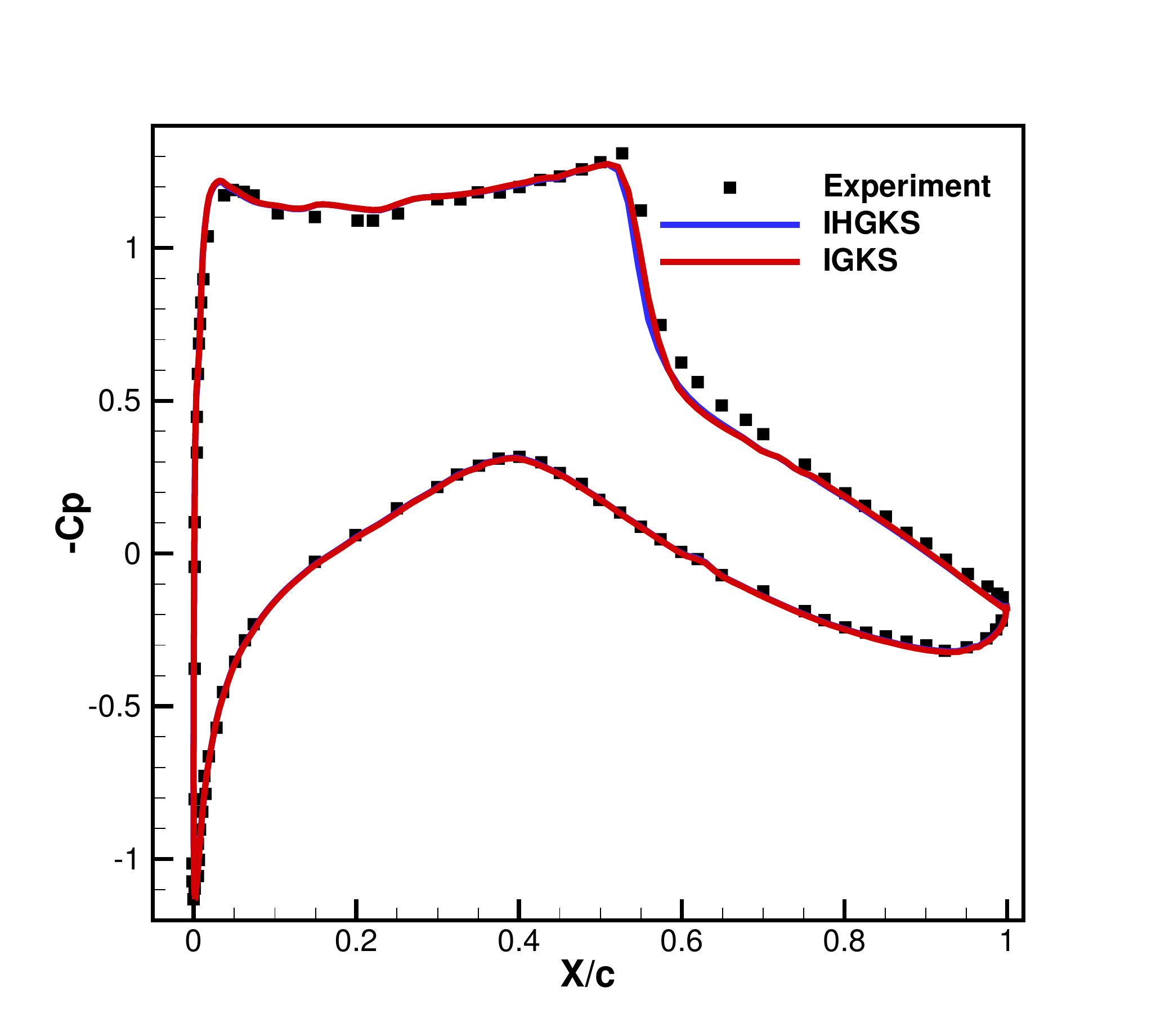}
	}
	\subfigure[]
	{
		\includegraphics[width=0.47\textwidth]{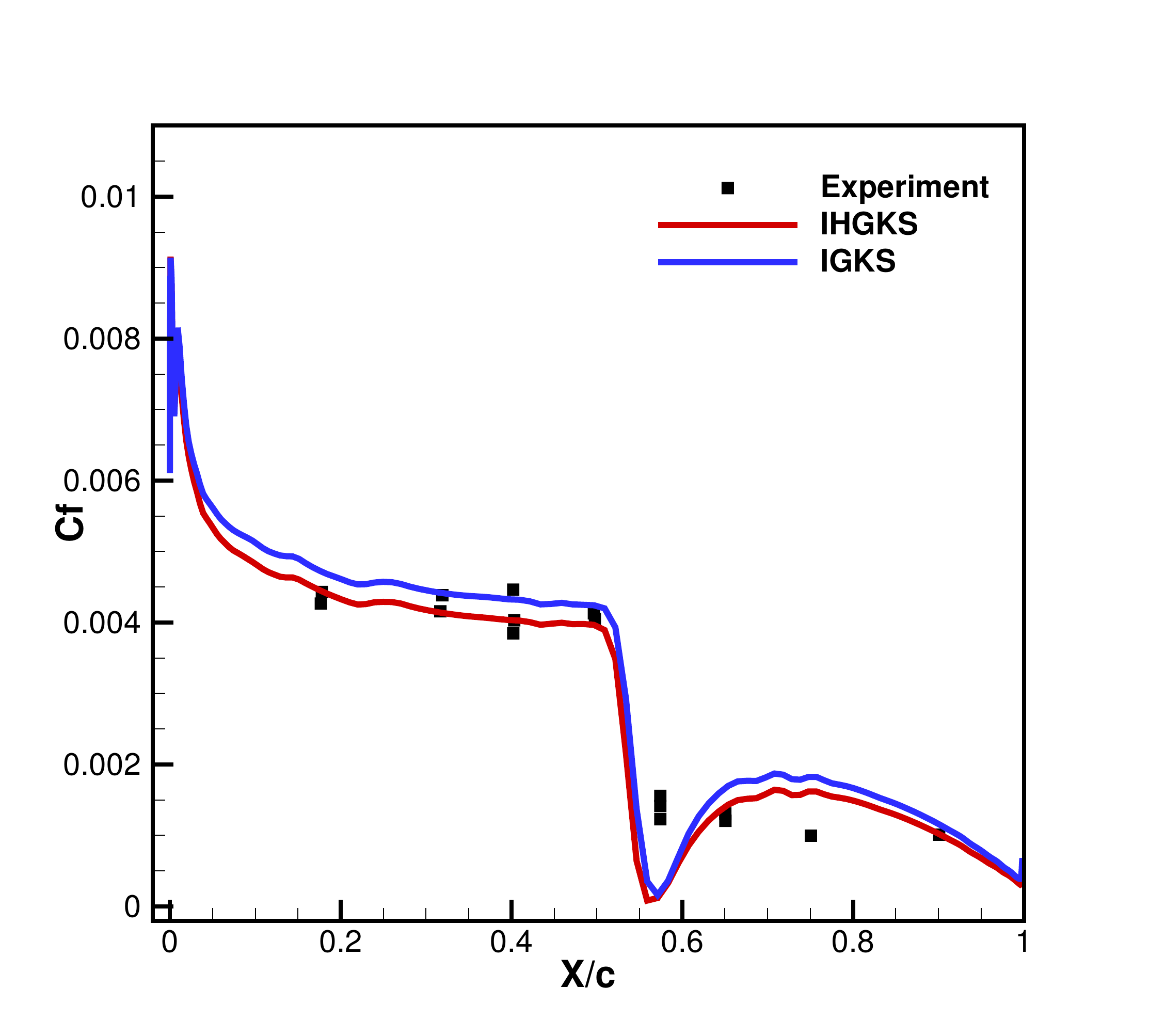}
	}
	\caption{\label{RAE2822_cp_cf}Pressure coefficient distributions $Cp$ around airfoil (a) and skin friction coefficient distributions $C_f$ on upper airfoil (b) from the experiment data, the current IHGKS, and the second-order IGKS.}
\end{figure}

CFL number is adopted as $CFL = 2$ and the total residual reduce down to the $6$ orders of magnitude. In Figure \ref{RAE2822_ma_miut}, Mach number contour and contour of normalized viscosity $\mu_t/\mu$ are presented based on the IHGKS. For this kind of transonic flows, the full flux as Eq.(\ref{formalsolution_neq}) is necessary for shock-capturing. Figure \ref{RAE2822_ma_miut} (a) shows the shock and its interaction with the turbulent boundary layer, which verify the robustness of the scheme on the capturing of shock. In Figure \ref{RAE2822_ma_miut} (b), the maximum eddy viscosity region is located near the trailing edge, which is different with the turbulence around NACA0012 airfoil with high angle of attack in Figure \ref{NACA0012_r3_cp_cf15} (b). To compare numerical solution from IHGKS/IGKS quantitatively, pressure coefficient distributions around airfoil and skin friction coefficient distributions on upper airfoil are presented in Figure \ref{RAE2822_cp_cf}. As presented in Figure \ref{RAE2822_cp_cf}, both the pressure coefficients and skin friction coefficients with the IHGKS and second-order IGKS match the experimental data from Cook et al. \cite{cook1979rae2822} well. This indicates turbulence model error dominates in this transonic flows instead of numerical discretization error. To our expectation, Figure \ref{RAE2822_cp_cf} (b) shows that skin friction coefficient with the IHGKS is slightly closer with the experiment data, compared with those using second-order IGKS. In view of the robustness of shock-capturing and  better friction coefficient provided by the IHGKS, the high-order scheme is still preferred to predict accurate shock-boundary interaction turbulence.

\subsection{RANS 3D case: transonic ARA M100 wing-body turbulence}
\begin{figure}[htp]
	\centering
	\subfigure[]
	{
		\includegraphics[width=0.47\textwidth]{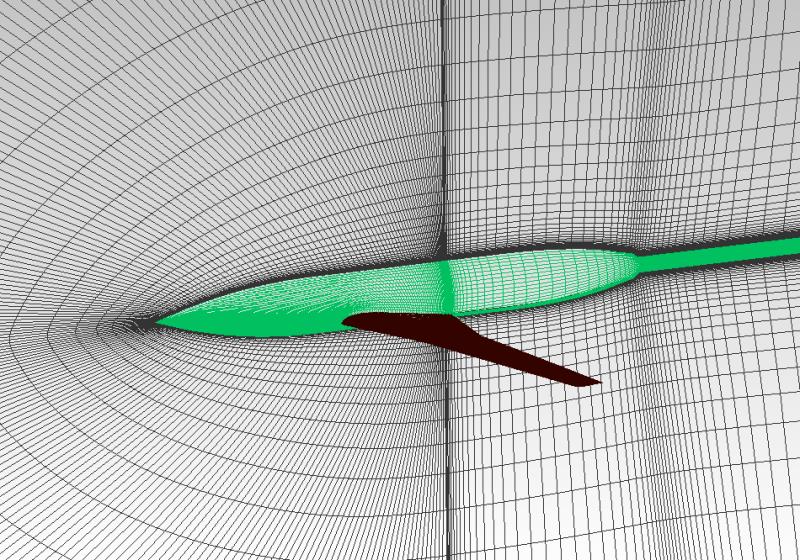}
	}
	\subfigure[]
	{
		\includegraphics[width=0.47\textwidth]{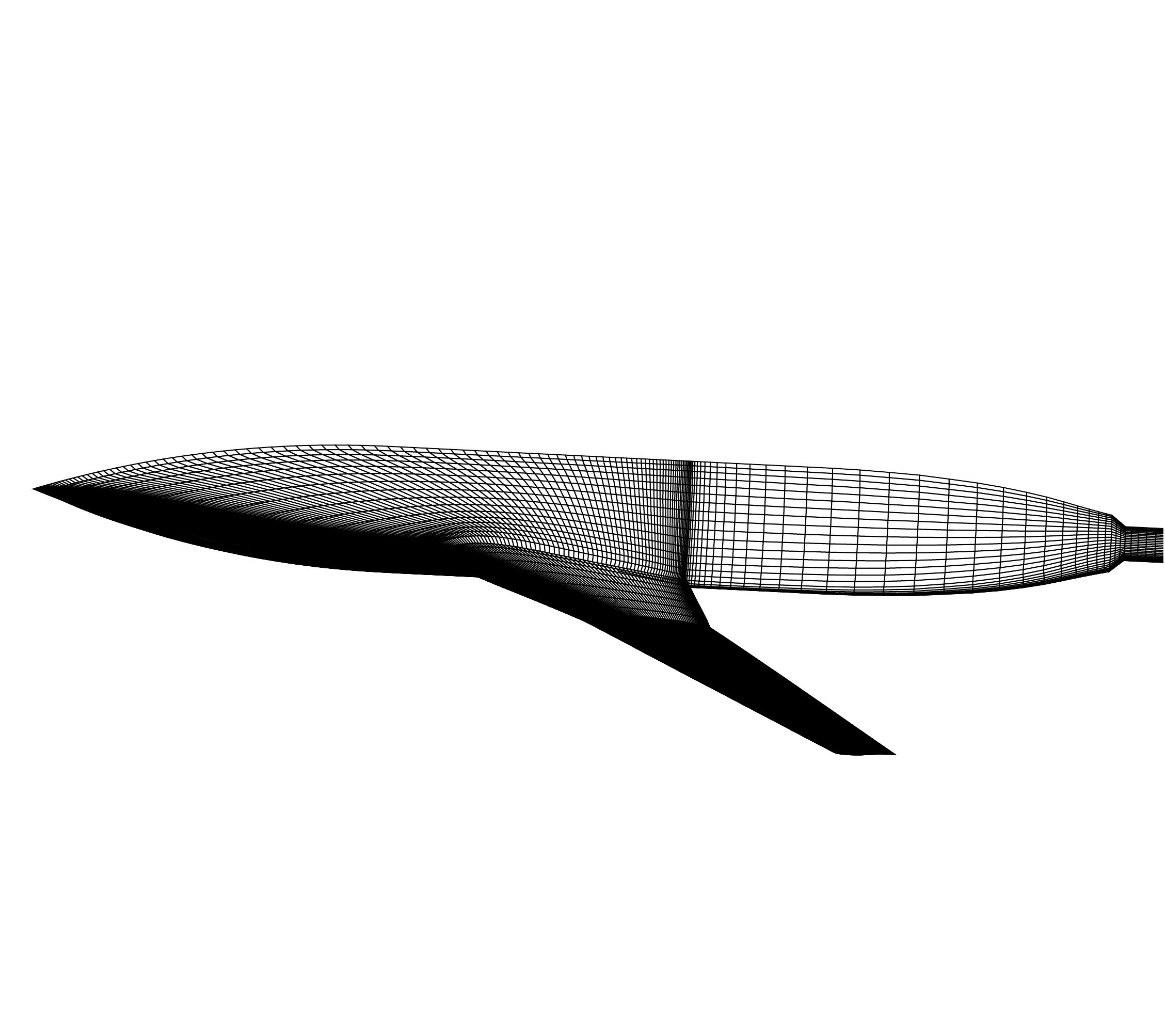}
	}
	\caption{Configuration of ARA M100 wing-body (a) and illustration of surface grid (b).}
	\label{M100_grid}
\end{figure}

Three-dimensional transonic turbulence around complex configuration of ARA M100 wing-body is simulated. This case is adopted to keep studying the robustness of capturing shock and validate the ability to simulate the three-dimensional real engineering turbulence by the current IHGKS. Typical cruising condition of  ARA M100 is the one corresponding to an angle of attack $\alpha = 2.873^{\circ}$, Mach number $Ma = 0.8027$, and a local chord based Reynolds number of $Re_{lc} = 1.31 \times 10^7$ (local chord $lc = 0.245$). In this paper, the computational domain, the boundary conditions, and the C-O type grid of $321 \times 57 \times 49$ provided by CFL3D Version 6 website \cite{CFL3D} are used, with an off wall $Y^{+}$ distribution as follows: $Y^+_{wing}=0.8$, $0.1 \leq Y^+_{fusel}  \leq 30$. Configuration  of ARA M100 wing-body and surface grid are shown in Figure \ref{M100_grid}, whose black part is the wing and the green part represents fuselage.

\begin{figure}[htp]
	\centering
	\subfigure[]
	{
		\includegraphics[width=0.43\textwidth]{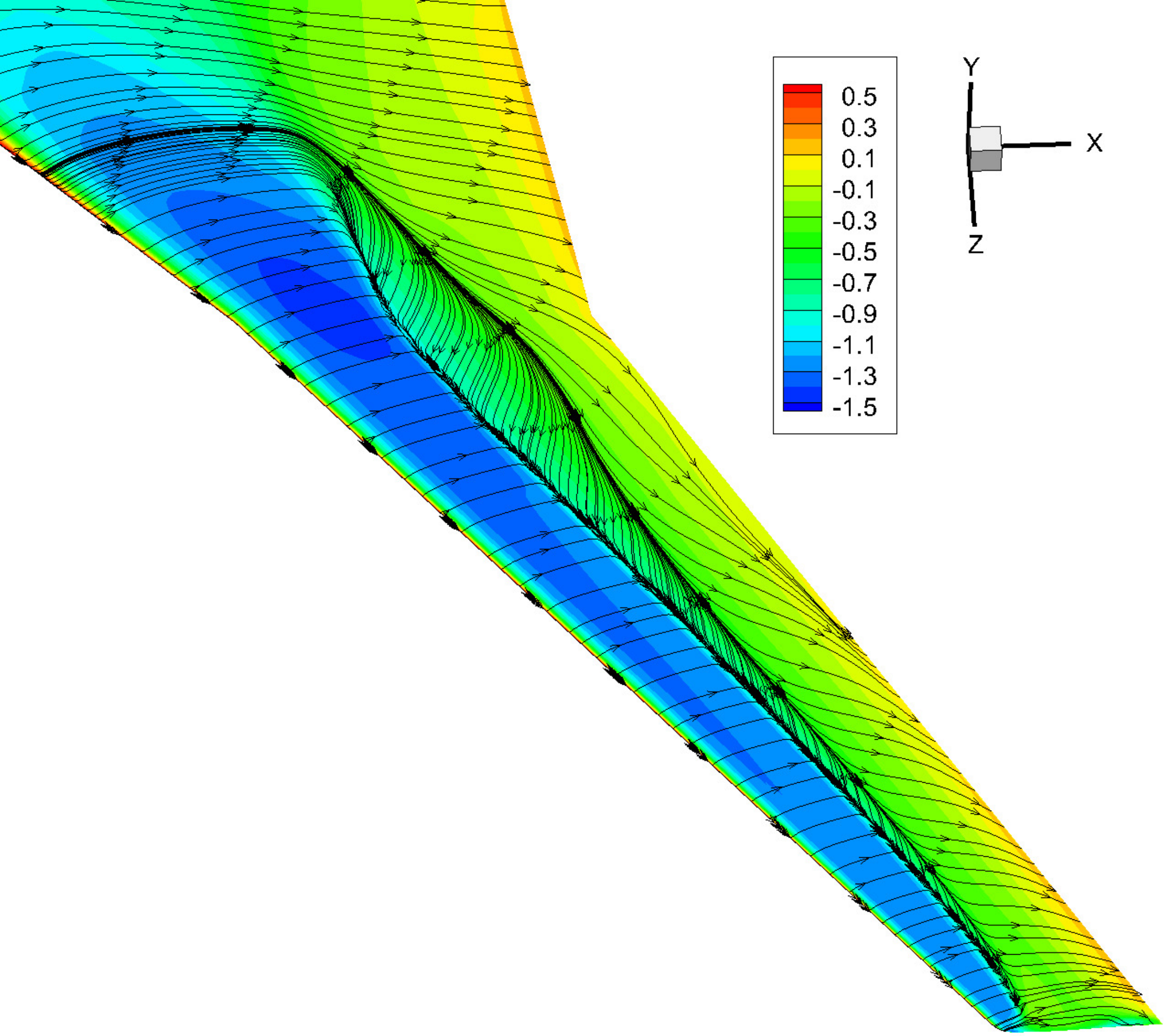}
	}
	\subfigure[]
	{
		\includegraphics[width=0.43\textwidth]{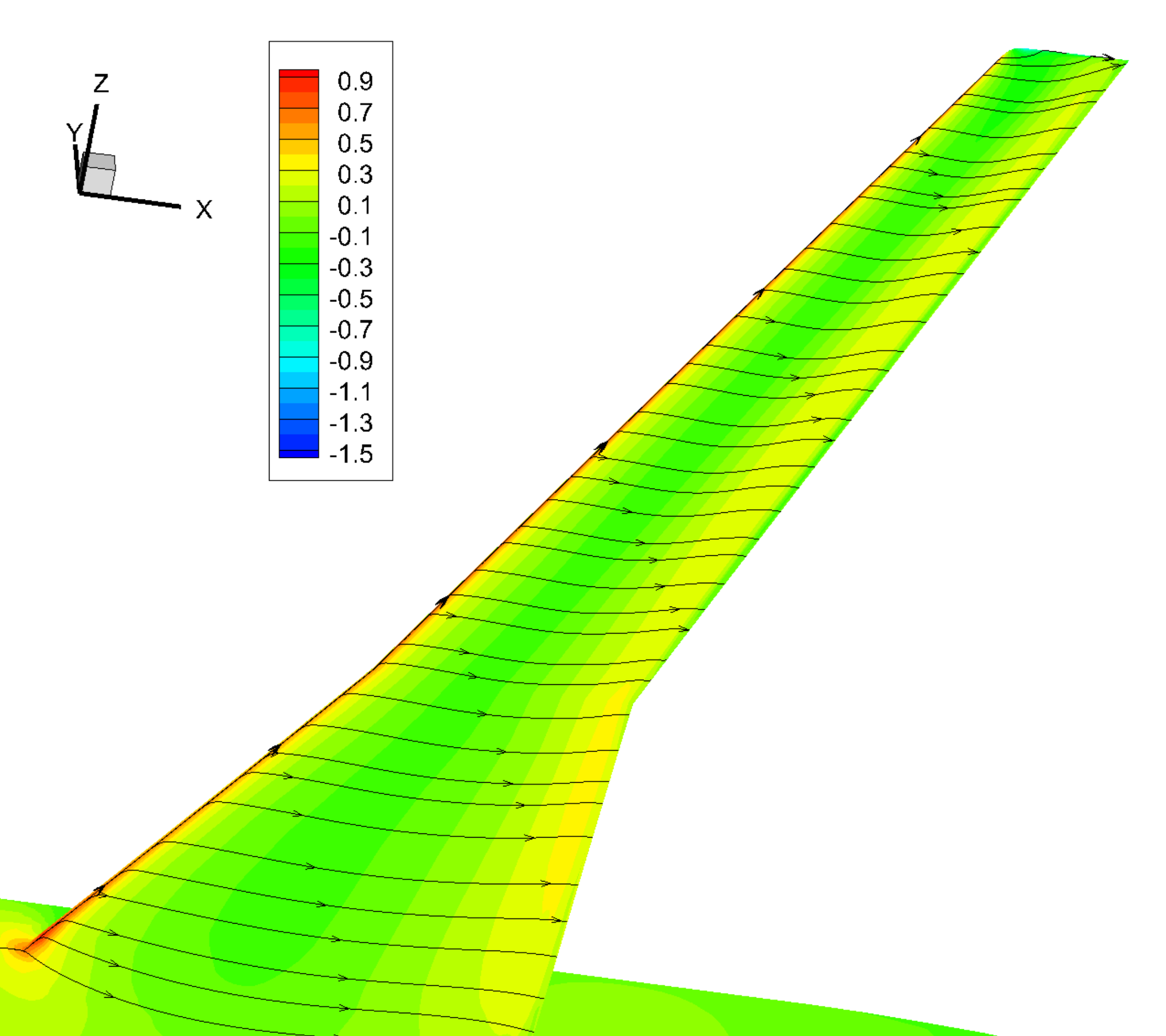}
	}
	\caption{Streamlines on the upper wing surface (a) and lower wing surface (b) of ARA M100 wing-body with pressure coefficient contours.}
	\label{M100_Cp_bubble}
\end{figure}

\begin{figure}[htp]
	\centering
	\subfigure[]
	{
		\includegraphics[width=0.47\textwidth]{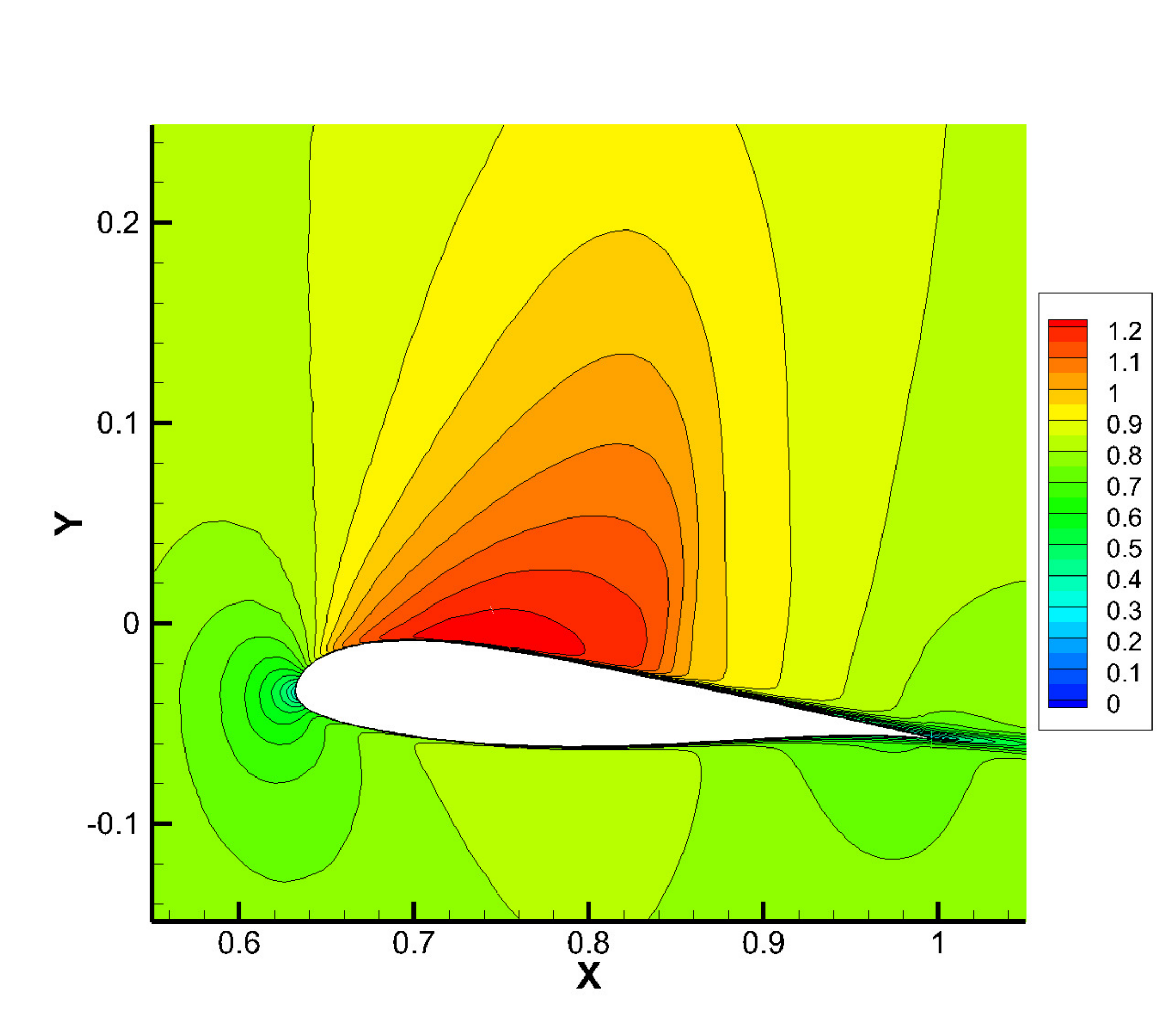}
	}
	\subfigure[]
	{
		\includegraphics[width=0.47\textwidth]{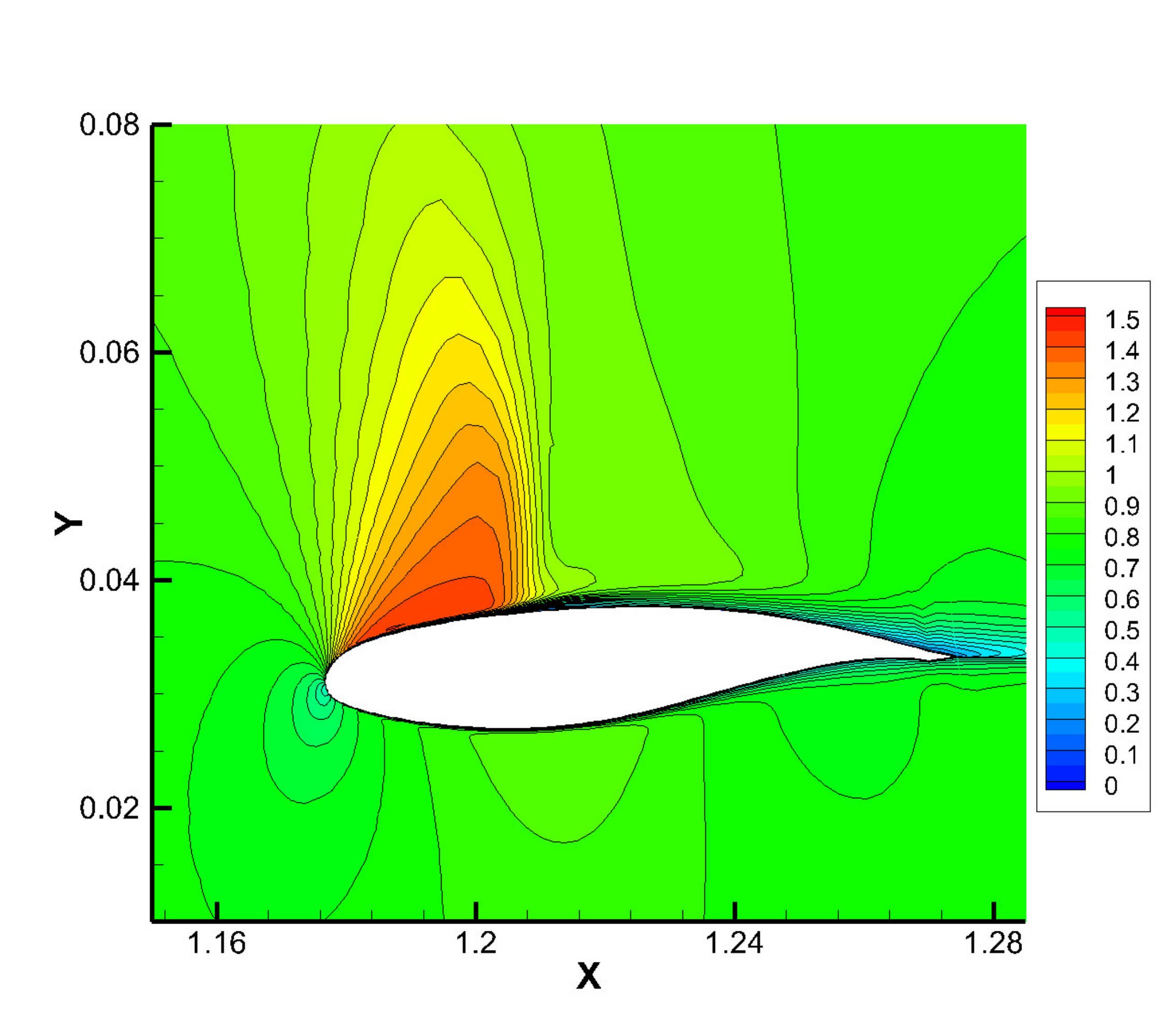}
	}
	\caption{Contours of Mach number at slice $Z/b = 0.019$ (a) and slice $Z/b = 0.935$ (b). $Z$ is the distance to the root chord plane and $b$ is the wing span of  ARA M100 wing-body.}
	\label{M100_ma}
\end{figure}

\begin{figure}[htp]
	\centering
	\subfigure[]
	{
		\includegraphics[width=0.47\textwidth]{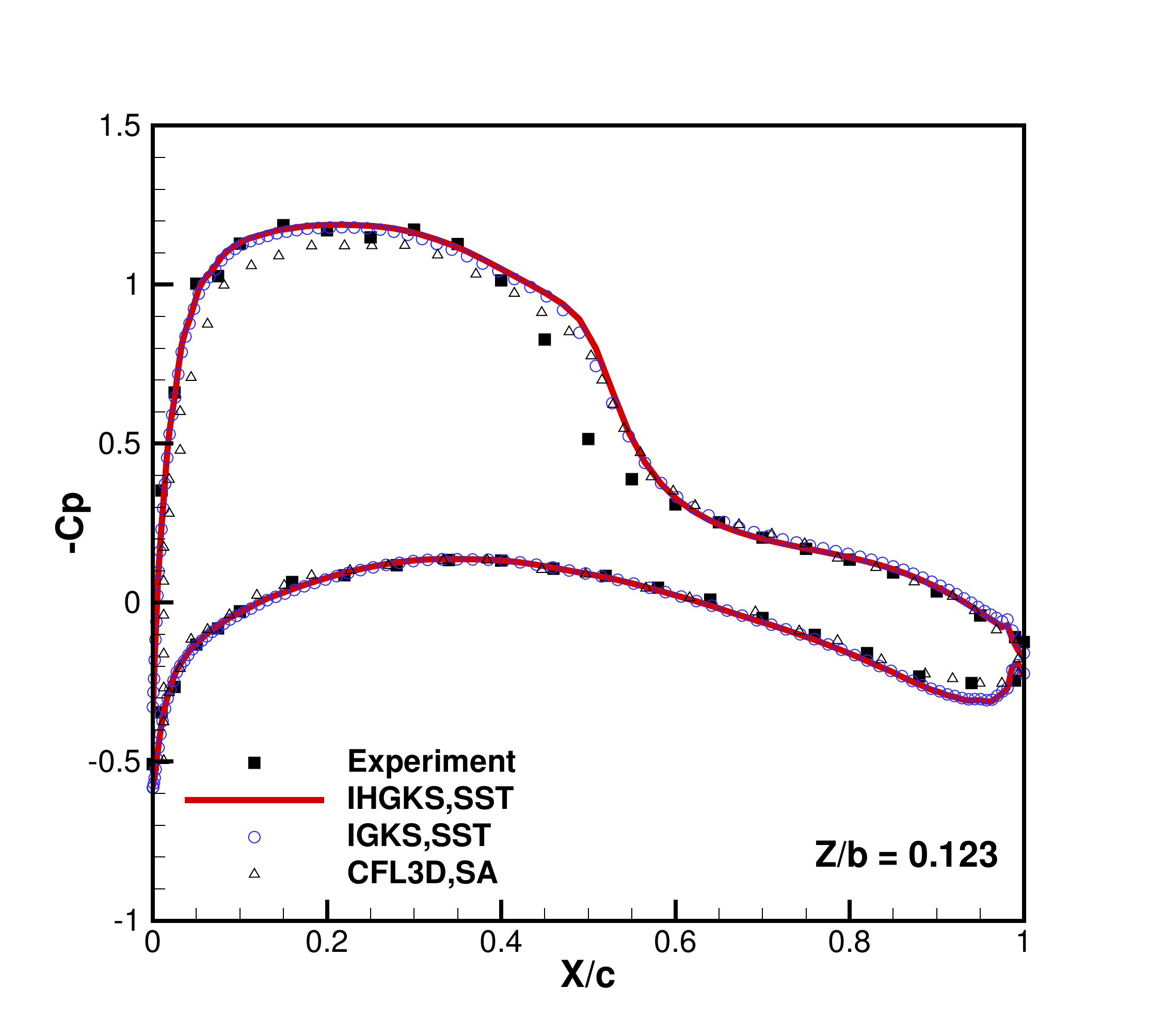}
	}
	\subfigure[]
	{
		\includegraphics[width=0.47\textwidth]{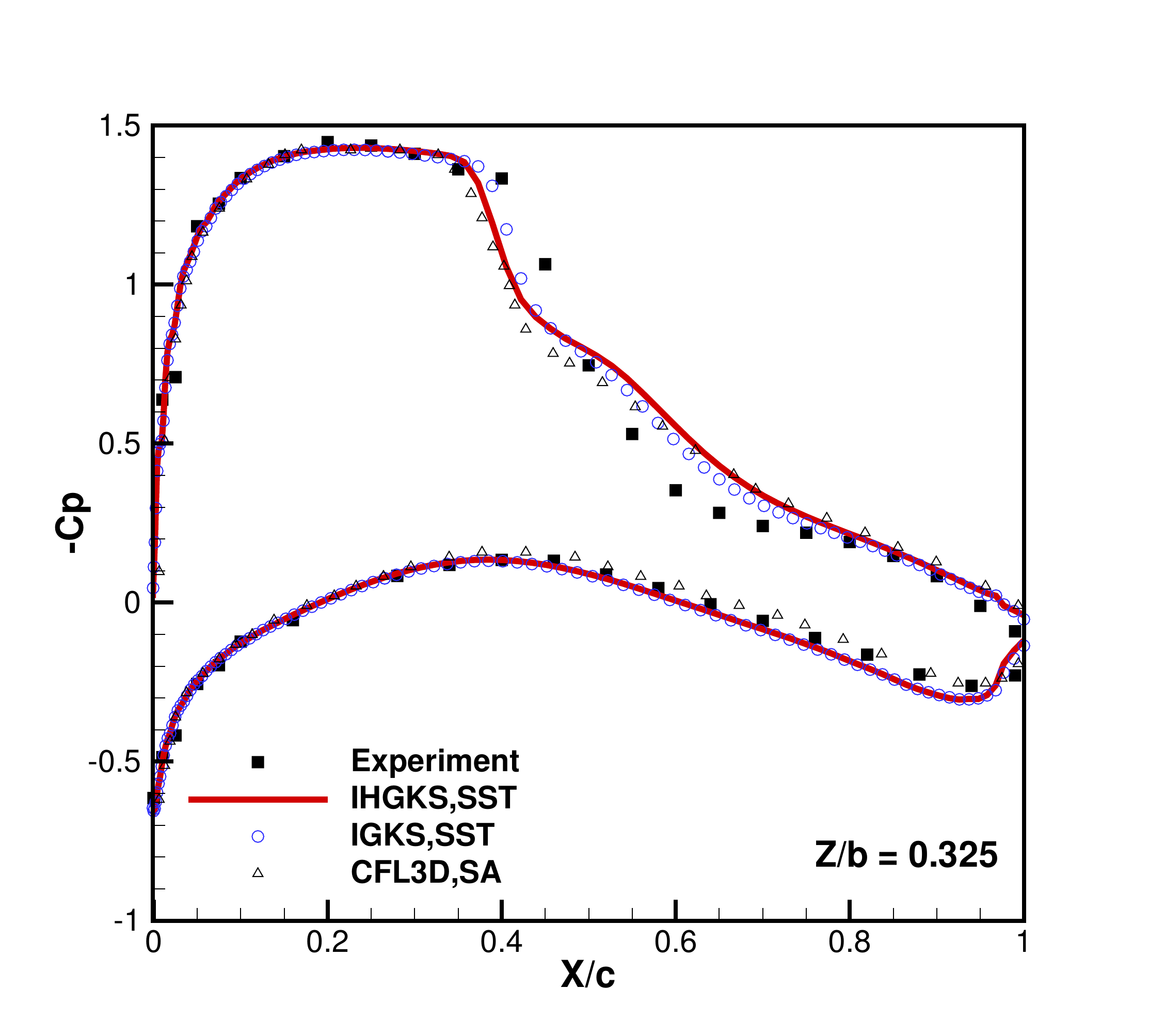}
	}
	\caption{Comparisons of pressure coefficient $Cp$ profiles at selected span-wise locations of ARA M100 wing-body from the experiment data, the current IHGKS, the second-order IGKS, and the second-order CFL3D. $c$ is the local chord length.}
	\label{M100_cp_1_3}
\end{figure}

The maximum CFL number of the IHGKS is $CFL = 1.8$, while the CFL number of the explicit HGKS is limited by $CFL = 0.25$. For this complex three-dimensional transonic turbulent flow, the total residual reduces down to $4$ orders of magnitude, with the full flux as Eq.(\ref{formalsolution_neq}). Figure \ref{M100_Cp_bubble} shows streamlines on the upper wing surface and lower wing surface with pressure coefficient contours. As shown in Figure \ref{M100_Cp_bubble} (a), the negative pressure coefficient regime is followed by a reverse flow regime, involving adverse pressure gradients. The reverse flow regime is enclosed by separated streamline and reattached line on the wing's suction side, and no reverse flow regime appears in the lower wing surface as  Figure \ref{M100_Cp_bubble} (b). Mach number contours of one slice $Z/b = 0.019$ near the root chord plane and the slice $Z/b = 0.935$ near the wing's tip are presented in Figure \ref{M100_ma}. These wing slices show the shock and its interaction with the turbulent boundary layer, which confirms the robustness of current scheme on the capturing of shock. The shock-boundary interaction is similar as above transonic RAE2822 turbulence in Figure \ref{RAE2822_ma_miut} (a). Comparisons of pressure coefficient $Cp$ profiles at two selected wing sections among the experimental data from the CFL3D website, the current IHGKS, the second-order IGKS, and results from CFL3D based on S-A model \cite{spalart1992one}, are plotted in Figure \ref{M100_cp_1_3}. These two selected wing sections cross the reverse flow regime. As the pressure coefficient $Cp$ from all schemes agree well with the experiment data, it confirms that both $k - \omega$ SST model and S-A model  have the ability to predict pressure-induced separation. For different turbulence models, numerical results show that S-A model is a little better than $k - \omega$ SST model on lower wing surface, while $k - \omega$ SST model outweighs S-A model on the upper wing surface. In Figure \ref{M100_cp_1_3}, compared with the obvious difference between different turbulence model, the current IHGKS almost takes no advantage than the second-order IGKS. It is not surprising as the turbulence model error dominates in this transonic three-dimensional complex RANS simulation rather than the numerical discretization error. This indicates that developing appropriate turbulence model is still the most important task for three-dimensional complex RANS simulation. For transition flows \cite{lee2001formation,lee2008transition}, the  turbulent model may play an even more important role.

\section{Conclusion}
In present work, targeting on accurate and efficient simulation of three-dimensional turbulent flows,
an implicit high-order GKS with LU-SGS method is developed under the two-stage fourth-order framework.
Vreman-type LES model for large eddy simulation and $k - \omega$ SST model for RANS simulation are coupled with the current IHGKS. The cases of incompressible decaying homogeneous isotropic turbulence, incompressible high-Reynolds number flat plate turbulent flow, incompressible turbulence around NACA0012 airfoil, transonic turbulence around RAE2822 airfoil, and transonic high-Reynolds number ARA M100 wing-body turbulence, are tested. The IHGKS shows the higher accuracy in space and time than that of the second-order IGKS, especially for smooth flows, obtaining more accurate turbulent flow fields on coarse grids. Compared with the explicit HGKS, the IHGKS provides great improvement on the computational efficiency. In addition, the robustness of the current IHGKS and the ability to capture shock are validated in the transonic two-dimensional and three-dimensional complex RANS simulation. Transonic cases indicate that turbulence model plays a leading role in the capturing of shock-boundary interaction turbulence. Developing appropriate turbulence model is still the most important task for complex turbulence simulation.

\section*{Conflict of interest statement}
None declared.

\section*{Acknowledgement}
Special thanks to Shuang Tan and Hualin Liu who provide the helpful discussion and suggestions.  The current research was supported by The National Key Research and Development Project of China (NO.2017YFB0202702), Hong Kong research grant council (16206617) and  National Science Foundation of China (11772281,91852114).

\section*{References}
\bibliographystyle{unsrt}
\bibliography{caogybib}

\end{document}